\documentclass[12pt,a4paper]{article}

\usepackage[utf8]{inputenc}
\usepackage[english]{babel}
\usepackage{amsfonts}
\usepackage{dmstyle}
\usepackage{cite}
\usepackage[normalem]{ulem}

\def \varsig {\varsigma}
\def \A {{\cal A}}

\begin{document}

\title{Uniqueness of extremal isolated horizons and their identification with horizons of all type D black holes}

\author{David Matejov\footnote{This work is dedicated to the memory of our late colleague Dr.~Martin Scholtz (1984--2019), the supervisor of~D.~M.. Unfortunately, due to the formal reasons he cannot be listed as its co-author.} 
       \quad and \quad
        Ji\v{r}\'{\i} Podolsk\'{y} 	\\[4mm]
	{\small Institute of Theoretical Physics, Faculty of Mathematics and Physics,} \\
	{\small Charles University, V~Hole\v{s}ovi\v{c}k\'ach~2, 180~00 Prague 8, Czech Republic
	} \\[3mm]
    {\small  E-mail: \texttt{d.matejov@gmail.com, podolsky@mbox.troja.mff.cuni.cz}}\\}

\maketitle

\begin{abstract}
We systematically investigate axisymmetric extremal isolated horizons (EIHs) defined by vanishing surface gravity, corresponding to zero temperature. In the first part, using the Newman-Penrose and GHP formalism we derive the most general metric function for such EIHs in the Einstein-Maxwell theory, which complements the previous result of Lewandowski and Pawlowski. We prove that it depends on 5 independent parameters, namely deficit angles on the north and south poles of a~spherical-like section of the horizon, its radius (area), and total electric and magnetic charges of the black hole. The deficit angles and both charges can be separately set to zero. In the second part of our paper, we identify this general axially symmetric solution for EIH with extremal horizons in exact electrovacuum Pleba\'{n}ski-Demia\'{n}ski spacetimes, using the convenient parametrization of this family by Griffiths and Podolsk\'y. They represent all (double aligned) black holes of algebraic type~D without a cosmological constant. Apart from a conicity, they depend on 6 physical parameters (mass, Kerr-like rotation, NUT parameter, acceleration, electric and magnetic charges) constrained by the extremality condition. We were able to determine their relation to the EIH geometrical parameters. This explicit identification of type D extremal black holes with a unique form of EIH includes several interesting subclasses, such as accelerating extremely charged Reissner-Nordstr\"om black hole (C-metric), extremal accelerating Kerr-Newman, accelerating Kerr-NUT, or non-accelerating Kerr-Newman-NUT black holes.
\end{abstract}

\vfil\noindent
PACS class:  04.20.Jb, 04.40.Nr, 04.70.Bw, 04.70.Dy, 04.20.-q, 97.60.Lf



\newpage
\tableofcontents

\newpage
\section{Introduction}

The Kerr(-Newman) solution \cite{Kerr1963, Newman1965} is a standard astrophysical model for rotating (and possibly charged) black holes. Despite several simplifying assumptions (stationarity of the full spacetime, asymptotic flatness, absence of matter
outside the black hole etc.), it is an excellent approximation for realistic black holes. Although real black holes are typically surrounded by accreting matter and external electromagnetic fields, these can be usually treated as a test matter with negligible backreaction on the geometry. In fact, the presence of the accretion disk is important in order to measure specific characteristics of the black hole, namely its spin, e.g.\ by the iron line method \cite{Reis}. Test-matter approach on the Kerr background has been also successful in explaining jets from active galactic nuclei via the Blandford-Znajek process \cite{Blandford-Znajek}.

Yet, there are good reasons to study black holes surrounded by matter beyond the test-field approximation. Kerr black holes comply with the no-hair theorem according to which a black hole in Einstein's general relativity is fully characterized by its mass, charge and angular momentum. This important theoretical result can be tested experimentally by measuring asymptotic multipole moments of the black hole, but it is expected that the presence of a matter will induce additional multipole moments. It was shown in \cite{Gurlebeck2015} that in the case of static black holes, contributions to multipole moments from the accreting matter can be disentangled from the contribution of the black hole and the moments coming from the black hole coincide with those of Schwarzschild -- in this sense, the no-hair theorem holds also for distorted static black holes. The situation for stationary case is not yet clear. However, experiments like Event Horizon Telescope \cite{Psaltis2016} have been proposed that will test the no-hair theorem and will be sensitive even to small deviations caused by surrounding matter. Hence, it is necessary to understand theoretically whether possible deviations from the theorem should be attributed to some alternative theory of gravity, or they are purely general relativistic effects.

In addition, the concept of the event horizon of a black hole is very rigid and has a teleological nature, meaning that it can be identified only after the full spacetime is known. That is a consequence of its global character: the event horizon
is a boundary of causal past of future null infinity. Our universe is not asymptotically flat and, moreover, it is desirable to have a good definition of a black hole which does not rely on the causal structure of the full spacetime, with its horizon identified locally or quasi-locally. Furthermore, there are more fundamental questions arising from the discovery of the Bekenstein-Hawking entropy of a black hole which is proportional
to its area. Entropy in thermodynamics is related to the number of possible microstates but it is far from being clear what these microstates should be in the case of a black hole, which is one of the main issues of the string theory or the loop quantum gravity.

Motivated by all these reasons, the concept of an \emph{isolated horizon} has been introduced, see \cite{Ashtekar2004-review} for an extensive review. These horizons are defined quasi-locally as null hypersurfaces with certain geometric properties (see Sec.~\ref{sec:isolated horizons}) but they do not restrict the spacetime --- they do not a priori require asymptotic flatness, and they admit the presence of radiation or other external matter. Hence, they can be much more realistic than the original Kerr solution.

Isolated horizons have played a significant role in the loop quantum gravity studies, but they have also found many useful applications in purely classical general relativity. For example, using this formalism it was recently possible to analyse, in full generality, the Meissner effect for extremal horizons \cite{Scholtz2017, Scholtz2018}.

\emph{Extremal} horizon represents a limit state of a black hole, which increases its charge, rotation, or another parameter to such an extreme degree that the horizon degenerates --- typically via a ``coalescence'' of two initially distinct horizons. Notable and well-known exact spacetimes of this type are extreme Reissner-Nordstr\"om, Kerr, Kerr-Newman, or Schwarzschild-de~Sitter black holes, see the corresponding sections in Chapters~9 and~11 of \cite{Griffiths2009}. In this limit state, the surface gravity of the horizon vanishes. Hence, according to black hole thermodynamics, such a black hole has zero temperature and does not radiate \cite{Bardeen1973}. From the quantum gravity perspective, this scenario is expected to be simpler to handle than the fully general case. For this reason, a considerable effort has been put into investigation of extremal black holes, including the near-horizon limits and classification of all possible such geometries --- see \cite{KunduriLucietti} for a review and references (interestingly, these near-horizon geometries belong to the Kundt family of nonexpanding metrics \cite{PodolskySvarc2013a, PodolskySvarc2013b}). In fact, extremality proved to be crucial in string theory calculations of the semi-classical formula for black hole entropy.

The extremal horizons have another distinguished property that --- despite possible distortions caused by an external matter --- assuming regularity, their intrinsic geometry is always locally isometric to the Kerr-Newman black hole, as the remarkable theorem by Lewandowski and Pawlowski shows \cite{Lewandowski-Pawlowski}. In fact, here we generalize this theorem to a much wider class of solutions, when the black holes are allowed to be penetrated by cosmic strings or struts. This is manifested by the presence of 1-dimensional topological defects extending from the poles of black hole horizon as conical singularities that physically represent cosmic strings or struts, which cause acceleration of the black holes. Although analogous results have been obtained and to some extent studied in previous literature \cite{Lewandowski-Pawlowski, KunduriLucietti, KunduriLucietti2009a, KunduriLucietti2009b, LiLucietti2013, Hajicek, Amsel2009} by various approaches, our systematic derivation using the Newman-Penrose formalism provides an independent insight into geometrical properties of such horizons. Moreover, we offer a physical interpretation of the result in its full generality, not investigated before.

This paper is organized as follows. In Sec.~\ref{sec:isolated horizons}, we review the necessary notation and basic definitions concerning the concept of isolated horizons. In Sec.~\ref{sec:extremal isolated horizons} we specialize on extremal isolated horizons, and we explicitly solve the constraint equations for a function describing the horizon geometry. In Sec.~\ref{sec:exact type D} we derive analogous result for the most general type~D black hole in a family of exact spacetimes of Pleba\'{n}ski-Demia\'{n}ski class. The last Sec.~\ref{sec:comparison} is dedicated to identification of the structure of these two solutions and mutual relations. Appendix~\ref{chap:NPFormalism} contains a summary of the NP and GPH formalisms.

\section{Isolated horizons}
\label{sec:isolated horizons}

Isolated horizons represent a mathematical framework for describing black holes that are in equilibrium with their neighborhood. They are quasi-local generalizations of globally defined event horizons. Our definitions here follow that of Ashtekar and Krishnan \cite{Ashtekar2004-review}.

\begin{definition} \label{def:NH}
A sub-manifold ${\NH \subset \man}$ of a spacetime $(\man,g_{ab})$ is said to be a \emph{non-expanding} horizon if the following conditions are satisfied:
\begin{enumerate}
\item $\NH$ is a null hypersurface of topology $\Real \times \spaceS^2$,
\item every null normal $l^a$ has vanishing expansion on $\NH$,
\item Einstein equations are satisfied on $\NH$ and energy-momentum tensor ${T_{ab}}$ is such that for eve\-ry
  future normal vector $l^a$, the vector ${{T^a}_b\, l^b}$ is also future pointing.
\end{enumerate}
\end{definition}

The first condition implies that the horizon can be foliated by topological 2-spheres interpreted as slices of constant time. The null normal $l^a$ is tangent to the generators of the horizon and is necessarily geodesic although not affinely parametrized. Its acceleration is the \emph{surface gravity} $\surg$ defined\footnote{For review of the Newman-Penrose formalism and notation, see appendix \ref{chap:NPFormalism}.} via $D l_a = \surg \, l_a$. Vanishing of the expansion of $l^a$ means that the area of the horizon does not change in time. Together with the energy condition 3, this is equivalent to zero flux of a matter through the horizon. The choice of the normal $l^a$ is unique up to rescaling by an arbitrary function. It is convenient to fix it by the following restriction:

\begin{definition}
An equivalence class $[\cdot]$ of vector fields is defined as
\begin{align*}
[ v^a ] = \{ X^a, \exists \lambda \in \Real^+: X^a = \lambda v^a \}.
\end{align*}
A \emph{weakly isolated horizon (WIH)} is a pair $(\NH,[l^a])$, where $\ham$ is a non-expanding horizon and $[l^a]$ is the equivalence class of a chosen null normal $l^a$ of $\ham$ such that
\begin{equation}
[\pounds_l, \mathcal{D}_a] \, l^b \eqNH 0, \label{WIH}
\end{equation}
where $\eqNH$ denotes equality on the horizon.
\end{definition}

Here, $\pounds_l$ is the Lie derivative along $l^a$, while $\mathcal{D}_a$ is  the induced covariant derivative defined by ${X^a \mathcal{D}_a \eqNH X^a \nabla_a}$ for every $X^a$ tangent to $\ham$.\footnote{Namely, in NP formalism $\mathcal{D}_a \eqNH n_a D - m_a \bar{\delta} - \bar{m}_a \delta$, c.f.\ \eqref{NP:NablaDecomposition}.}

The normal $l^a$ can be completed to a Newman-Penrose (NP) \emph{null tetrad} $(l^a, n^a, m^a, \bar{m}^a)$ \cite{Stewart-1993}, where the spatial (complex) vectors $m^a, \bar{m}^a$ span a tangent space of a particular spherical section $\spaceS_0^2$ of $\ham$ and are propagated onto the whole horizon by the requirement
\begin{equation}
\pounds_l \, m^a \eqNH 0 \quad \Leftrightarrow \quad \bar{\varepsilon} \eqNH \varepsilon \eqNH \frac{1}{2}\, \surg, \label{eq:LieDragginOfm}
\end{equation}
where $\varepsilon$ is the NP spin coefficient. Fixing $l^a$ and $m^a$, the vector $n^a$ is then determined by the normalization conditions ${l_an^a=1}$ and ${m_an^a=0}$ of the NP tetrad.

The properties of thus defined weakly isolated horizon allow us to construct adapted coordinates $(v, x^1, x^2)$ on $\ham$ such that
\begin{align} \label{eq:AdaptedCoordinates}
l^a \eqNH \pd^a_v, \quad m^a \eqNH \xi^I (x^J)\, \pd^a_I, \quad\hbox{where}\quad I,J \in \{1,2\},
\end{align}
for suitably chosen functions $\xi^{I}$. In order to extend these coordinates also out of the horizon we send geodesics in the direction of $n^a$ from every point of $\ham$, i.e. ${\Delta n^a = 0}$. These geodesics might be affinely parametrized by a parameter $r$ such that $n^a = \bpd_r^a$. Then the remaining vectors and coordinates are propagated analogously,
\begin{align} \label{eq:PropagationOfCoordinates}
\Delta l^a = \Delta m^a = 0, \quad \Delta v = \Delta x^I = 0,
\end{align}
which in turn implies
\begin{align*}
\gamma = \tau = \nu = 0, \quad \alpha + \bar{\beta} = \pinp.
\end{align*}
On the horizon $\ham$, it also holds
\begin{align} \label{eq:spincoefonH}
\kappa \eqNH \rho \eqNH \sigma \eqNH 0, \quad \mu \eqNH \bar{\mu}, \quad \Psi_0 \eqNH \phi_0 \eqNH 0.
\end{align}
This means that, being non-expanding, non-twisting and shear-free, its geometry must belong to the Kundt class of geometries \cite{Stephani, Griffiths2009, PodolskySvarc2013a, PodolskySvarc2013b}.

Moreover, the following coefficients are time-independent in the sense of vanishing derivative along $l^a$,
\begin{align}
D \pinp \eqNH D \alpha \eqNH D \beta \eqNH D \eps \eqNH 0, \quad  D \Psi_2 \eqNH D \phi_1 \eqNH 0. \label{eq:TimeIndepencenceOnWHI}
\end{align}
The condition (\ref{WIH}) gives ${\delta \eps \eqNH 0}$ which, using \eqref{eq:LieDragginOfm},  implies that the \emph{surface gravity $\surg$ is constant across the horizon}. This result is also known as the zeroth law of black hole thermodynamics. The gauge freedom in the choice of $l^a$ implies that $\surg$ does not acquire a unique value for a given WIH, but  in our case that poses no problem since we will concentrate on the study of \emph{extremal horizons} for which unambiguously ${\surg = 0}$.

As we have already mentioned, WIH admits the presence of a matter or even radiation outside the horizon and therefore represent a black hole horizon much more generally than event horizons or the particular Kerr solution. Although WIHs characterize equilibrium situation by definition, they admit a certain degree of time dependence \cite{Krishnan-2012}. Unfortunately, this is still not fully suitable for our purpose. We need to introduce a stronger notion of an ``isolation'':

\begin{definition} \label{def:IH}
An \emph{isolated horizon (IH)} is a WIH such that for \emph{every} vector $X^a$  tangent to $\ham$ the following condition holds
\begin{equation}
[\pounds_l, \mathcal{D}_a] X^b \eqNH 0. \label{IH}
\end{equation}
\end{definition}

As a consequence of the definition (\ref{IH}) we also obtain
\begin{align}
D \lambda \eqNH D \mu \eqNH 0.   \label{eq:LamdbaMuTimeIndependence}
\end{align}
Hence, all the initial data on $\mathcal{H}$ are time-independent. While the condition \eqref{WIH} for WIH is merely a gauge fixing \cite{Scholtz2017}, the condition \eqref{IH} for IH puts the direct restriction on admitted geometry.

Although  IH is naturally imposed in the context of stationarity, it might be too restrictive within the current setting in particular applications.
For instance, in order to prove the Meissner effect in a greater generality, the definition of the IH had to be further generalized in \cite{Scholtz2018} to the concept of \emph{almost isolated horizon}. The authors showed that the whole construction is valid even under assumption of topology $\Real \times \mathcal{K}$, where $\mathcal{K}$ is a compact 2-manifold.\footnote{The definition of the almost isolated horizon again allows certain amount of time-dependence of the initial data. Instead of \eqref{eq:LamdbaMuTimeIndependence} one assumes only ${D \lambda \eqNH 0}$. Here, however, we need both conditions.}

For the purpose of this article, we will similarly assume  that $\mathcal{K}$ has topology of a 2-sphere \emph{possibly pierced by strings (or struts) on the two poles}, which produce \emph{deficit angles} $\delta_{\pm}$, respectively\footnote{We assume that $\mathcal{K}$ has the structure of a differentiable 2-manifold, except in small neighbourhoods around its poles which do not have this structure because of the conical singularities. Then $\mathcal{K}$ is not compact, just bounded. However, this rather technical issue does not affect our investigation.}.
 In addition, we will impose the axial symmetry in the sense of \cite{Scholtz2018}, which is a generalization of the approach of \cite{Ashtekar2004-multipolemoments}:

\begin{definition}
A horizon section $\mathcal{K}$ with the topology $\spaceS^{\delta_+}_{\delta_-}$ equipped with a spatial metric $q_{ab}$ is said to be axially symmetric if there exists a Killing vector field $\phi^a$ with closed orbits, which vanishes exactly at two points of $\mathcal{K}$. These points are called \emph{poles}.
\end{definition}

Such a horizon section $\mathcal{K}$ can be coordinatized by two functions $\zeta$ and $\phi$, which take the values ${\zeta \in [-1,1]}$ and ${\phi \in [0, 2\pi)}$. In these coordinates, the induced (negative definite) metric $q_{ab}$ on the sections of $\mathcal{H}$ of constant time $v$ acquires the canonical form
\begin{align}  \label{CanonicalMetric}
q_{ab}\, \dd x^a \dd x^b \equiv - R^2\,\Big( \,\frac{1}{f(\zeta)}\, \dd \zeta^2 + f(\zeta)\,\dd \phi^2 \,\Big),
\end{align}
where $R$ is ``Euclidean'' radius defined by the area $A$ of $\mathcal{K}$ via the relation ${4\pi R^2 = A}$. Because of condition \eqref{eq:LieDragginOfm}, this metric is Lie-constant along $l^a$.\footnote{In NP formalism the metric of $\ham$ is $q_{ab} = -m_a \bar{m}_b - m_b \bar{m}_a$. Notice the degeneracy of $q_{ab}$ following from the null character of $\ham$.}

In particular, for ${f=1-\zeta^2}$ with ${\zeta=\cos\theta}$, the metric (\ref{CanonicalMetric}) takes the standard form
\begin{align}  \label{CanonicalMetricSphere}
q_{ab}\, \dd x^a \dd x^b \equiv - R^2\,\big( \,\dd \theta^2 + \sin^2\theta\,\dd \phi^2 \,\big).
\end{align}
Consequently, ${f' = - 2\zeta=-2\cos\theta}$ which yields ${f'=-2}$ for ${\theta=0}$ and  ${f'=2}$ for ${\theta=\pi}$.

The function $f(\zeta)$ in (\ref{CanonicalMetric}) can be chosen arbitrarily, provided it satisfies the boundary conditions at the poles ${f(\pm 1) = 0}$. In \cite{Lewandowski-Pawlowski}, another condition was imposed, namely $f'(\pm 1) = \mp 2$ which makes the sphere ``elementary flat'' \cite{Stephani} in the sense that there are no conical singularities (deficit angles) around the poles at ${\zeta=\pm1}$. In order to \emph{relax the regularity in the above sense}, the values of derivatives of $f$ are prescribed in the more general form \cite{Scholtz2018}
\begin{align} \label{BoundaryConditions}
f'(\pm 1) = \mp \, 2 \Big( 1 + \frac{\delta_{\pm}}{2\pi}\Big),
\end{align}
where $\delta_{\pm}$ are the corresponding \emph{deficit angles at}  ${\zeta = \pm 1}$. Such topological defects have non-trivial effects on the spacetime and can be interpreted as \emph{cosmic strings (or struts) extending through} the horizon poles and causing acceleration of the black hole as in the $C$-metric, see e.g. Chapter~14 in \cite{Griffiths2009}.

A convenient choice of the spatial vector $m^a$ on $\ham$ is
\begin{align} \label{eq:VectorMonIH}
m^a \eqNH \frac{1}{\sqrt{2}\,R} \Big(\sqrt{f(\zeta)}\, \pd^a_{\zeta} + \frac{\ii}{\sqrt{f(\zeta)}}\, \pd^a_{\phi} \Big),
\end{align}
normalized as ${m_a \bar{m}^a=-1}$.
The only independent component of the connection on $\ham$ is then given by the coefficient $a$ defined by (\ref{NP:defa}),
\begin{align} \label{eq:SpinCoefficientA}
a \equiv m_a \bar{\delta}\, \bar{m}^a = \alpha - \bar{\beta} \eqNH - \frac{1}{2\sqrt{2}R} \,\frac{f'(\zeta)}{\sqrt{f(\zeta)}}.
\end{align}
With this choice, $a$ is real on the horizon,  $\bar{a} \eqNH a$, as well as the derivative operator $\delta \equiv m^a \nabla_a \eqNH \bar{\delta}$ acting on a scalar function, namely
\begin{align} \label{eq:OperatorDelta}
\delta \fii \eqNH \frac{1}{\sqrt{2}\,R} \sqrt{f(\zeta)}\, \pd_{\zeta} \fii,
\end{align}
for an arbitrary function $\fii = \fii(\zeta)$.

\section{Extremal isolated horizons (EIH)}
\label{sec:extremal isolated horizons}

In this section, we investigate isolated horizons, which are \emph{extremal}. Throughout the text, we denote such a horizon by a shorthand notation EIH.

Extremal horizons are characterized by \emph{vanishing surface gravity}, that is
\begin{align} \label{eq:extremality}
\surg = 0.
\end{align}
It puts a geometric restriction on the metric function $f(\zeta)$ in  (\ref{CanonicalMetric}), in particular via the Ricci identities, and reduces the dependence on the free data represented by the spin coefficients $\mu$ and $\lambda$. In fact, the solution can be found explicitly, and it is unique.

Following \cite{Scholtz2018}, in Sec.~\ref{sec:extremal isolated horizons1} we will proceed at first by solving the necessary constrains for the electromagnetic field and the spin coefficient $\pinp$. Then, in Sec.~\ref{sec:extremal isolated horizons2}  we will investigate the equation for the metric function $f(\zeta)$ and solve it.

\subsection{Electromagnetic field}
\label{sec:extremal isolated horizons1}

In this paper we are interested in (electro)vacuum spacetimes, \emph{assuming that the electromagnetic field $F_{ab}$ everywhere shares the axial symmetry and time-independence} with the gravitation field, see (\ref{eq:TimeIndepencenceOnWHI}), namely that on the horizon
\begin{align}
D \phi_2 \eqNH 0. \label{eq:elmagAssumption}
\end{align}
Since ${\phi_0 \eqNH 0}$ in general, see (\ref{eq:spincoefonH}), the relevant Maxwell equation \eqref{NP:Dphi2} is thus reduced to
\begin{align*}
\delta \phi_1 + 2\pinp\, \phi_1 - \surg \,\phi_2 \eqNH 0,
\end{align*}
with $\phi_0, \phi_1, \phi_2$ being the null tetrad components of the electromagnetic field tensor $F_{ab}$.

Furthermore, the spin coefficient $\pinp$ is time-independent on every (weakly) isolated horizon, c.f. equation \eqref{eq:TimeIndepencenceOnWHI}. Its values are constrained by the Ricci identity \eqref{np:RI:Dlambda}, which on the horizon reads
\begin{align*}
\bar{\eth} \pinp \eqNH \surg \,\lambda - \pinp^2,
\end{align*}
where $\bar{\eth}$ is the spin lowering operator \eqref{NP:Eth} (and $\eth$ its spin raising counterpart) \cite{Goldberg1967}.

By applying the condition (\ref{eq:extremality}) that the horizon is extreme, ${\surg = 0}$, these two equations reduce to
\begin{align}  \label{eq:EquationForPi}
\bar{\eth} \pinp \eqNH - \pinp^2,\qquad
\delta \phi_1 + 2\pinp\, \phi_1 \eqNH 0.
\end{align}
Using the coordinates introduced in the previous section, we obtain explicitly
\begin{align*}
\pd_{\zeta}\, \pinp - \frac{\pd_{\zeta}f}{2 f} \,\pinp + \frac{\sqrt{2} R}{\sqrt{f}}\,\pinp^2 \eqNH 0, \qquad
\pd_{\zeta}\, \phi_1 + \frac{2\sqrt{2}R}{\sqrt{f}} \,\pinp \, \phi_1 \eqNH 0.
\end{align*}
These equations have the following simple general solutions \cite{Scholtz2018}
\begin{align} \label{ExtremalPiAndPhi}
\pinp \eqNH \sqrt{\frac{f}{2}} \,\frac{1}{R\,(\zeta + c_{\pi})}, \qquad
\phi_1 \eqNH \frac{c_{\phi}}{(\zeta + c_{\pi})^2},
\end{align}
where $c_{\pi}$ and $c_{\phi}$ are complex integration constants.

The physical meaning of the constant $c_{\phi}$ can be determined using the Gauss law expressing the \emph{total electric and magnetic charges}, see \cite{Krishnan-2012}, as
\begin{align} \label{eq:IHcharge}
Q \equiv Q_E + \ii\, Q_M = \frac{1}{2 \pi} \oint_{\mathcal{K}} \! \phi_1 \; \dd \zeta \wedge \dd \phi
   = \frac{2 R^2}{c_{\pi}^2 - 1}\, c_{\phi}.
\end{align}
Inverting this relation gives us
\begin{align} \label{CPhi}
c_\phi &= \frac{Q}{ 2 R^2}\,(c_\pi^2 - 1).
\end{align}
It has a well-defined limit for ${Q \rightarrow 0}$. Vanishing of $c_{\phi}$ is then equivalent to zero electromagnetic charge $Q$, consistent with the vanishing electromagnetic field $\phi_1$ in \eqref{ExtremalPiAndPhi}. In this case, the spacetime is a vacuum solution.

Notice that the electromagnetic field component $\phi_1$ does not depend on the actual metric on $\ham$, but it depends on the topology encoded in the complex constant $c_{\pi}$, as will be shown below.

\subsection{The horizon geometry}
\label{sec:extremal isolated horizons2}

It has been previously shown that extremal isolated horizons EIH \emph{with regular axes} have the geometry, which is necessarily isometric to the intrinsic geometry of the Kerr-Newman black hole \cite{Lewandowski-Pawlowski}, see also \cite{KunduriLucietti, KunduriLucietti2009a, KunduriLucietti2009b, LiLucietti2013, Hajicek, Amsel2009} for analogous results using different methods. Here, we generalize the result of \cite{Lewandowski-Pawlowski} to a wider class of solutions by admitting topological defects interpreted as cosmic strings (or struts), which are essential part of many exact solutions of Einstein's equations, in particular the $C$-metric or the Taub-NUT solution \cite{Griffiths2009}.

In NP formalism, the spin coefficient $\pinp$ is subject to the Ricci identity \eqref{np:RI:Dlambda} which, in the extremal case, yields the explicit solution \eqref{ExtremalPiAndPhi}. Using the time-independence property of the isolated horizons \eqref{eq:LamdbaMuTimeIndependence} we notice that in the Ricci identity \eqref{np:RI:Dmu} the only undetermined NP quantity is the $\Psi_2$ component of the Weyl tensor, provided $\mu$ is a part of the free data. However, $\Psi_2$ has to satisfy \eqref{np:RI:deltaalpha}, and then the only unknown quantity in the equation remains the metric. Thus, using the extremality condition \eqref{eq:extremality}, we have explicitly
\begin{align} \label{EquationForMetricFunction}
\eth \pinp \eqNH  - \pinp \pinpb -\Psi_2, \qquad
\eth \pinp -\overline{\eth \pinp} \eqNH 2 \Psi_2 +2a^2 - 2 \delta a + 4|\phi_1|^2,
\end{align}
where the second equation is (complex) algebraic constraint for $\Psi_2$. Combining the imaginary parts of these equations does not contain any new information, it is always trivially satisfied. Taking their real parts, eliminating $\Psi_2$, and using equation \eqref{eq:EquationForPi} we arrive at
\begin{align*}
a^2 - \delta a + 2|\phi_1|^2 \eqNH {\textstyle \frac{1}{2}} (\pinp - \pinpb)^2 + a (\pinp + \pinpb).
\end{align*}
The coefficient $a$ is defined by \eqref{eq:SpinCoefficientA}, while the derivative operator $\delta$ is obtained from \eqref{eq:OperatorDelta}. Hence, in our choice of the coordinates of the metric (\ref{CanonicalMetric}) and using the previous results \eqref{ExtremalPiAndPhi}, this key equation takes the explicit form
\begin{align*}
|\zeta + c_{\pi}|^4\, f'' + (2\zeta + c_{\pi} + \bar{c}_{\pi}) |\zeta + c_{\pi}|^2\, f'
  - (c_{\pi} - \bar{c}_{\pi})^2\,f + 8 R^2 |c_\phi|^2 \eqNH 0,
\end{align*}
which is a specific differential equation for the metric function $f(\zeta)$. Its \emph{general} solution might be found explicitly. After integration we obtain
\begin{align}\label{MetricFunction}
f(\zeta) &= \frac{4|c_\phi|^2 R^2(1-\zeta^2)}{(|c_\pi|^2 - 1)\,|\zeta+c_\pi|^2},
\end{align}
where we have employed the boundary conditions ${f(\pm 1)=0}$ to fix the integration constants. The constant $c_{\phi}$ is uniquely related to the electromagnetic charge $Q$ via \eqref{CPhi}, while the value of $c_{\pi}$ can be found from our generalized regularity condition \eqref{BoundaryConditions} as
\begin{align*}
c_{\pi} = \frac{1}{4\pi + \delta_- + \delta_+ -4\pi  |Q|^2 R^{-2}}\left( \delta_- - \delta_+
\pm 2\, \ii\, \sqrt{(2\pi +\delta_-)(2\pi +\delta_+) - 4 \pi^2 |Q|^4 R^{-4}}\,\right).
\end{align*}
Two solutions are possible due to the symmetry ${c_{\pi} \leftrightarrow \bar{c}_{\pi}}$ in \eqref{MetricFunction}.
After substituting into \eqref{MetricFunction} and some algebraic manipulation, we arrive at our \emph{main result}:

\begin{theorem} \label{th:MetricIH}
Let $(\NH, [l^a])$ be an axially symmetric extremal isolated horizon (EIH) of topology $\spaceS^{\delta_+}_{\delta_-}$. Then the geometry of its spherical sections is described by an induced metric $q_{ab}$ in the form \eqref{CanonicalMetric}, where the \emph{dimensionless} metric function $f(\zeta)$ explicitly reads
\begin{align} \label{th:MetricFunction}
  f(\zeta) &= \frac{2}{\pi} \frac{(2\pi+\delta_-)(2\pi+\delta_+)(1-\zeta^2)}
  { 4\pi(1 + \zeta^2)+ \delta_-(1 + \zeta)^2 +\delta_+(1-\zeta)^2
  + 4\pi |Q|^2 R^{-2}(1-\zeta^2)}.
\end{align}
Moreover, $f(\zeta)$ is unique and depends on 5 independent parameters, namely
${\delta_-, \delta_+, R}$ and ${Q \equiv Q_E + \ii\, Q_M}$. It is well-behaved, and any of these parameters (except $R$ when ${|Q|\neq 0}$) can be set to zero.
\end{theorem}

\emph{Without charges} ${(Q_E =0= Q_M)}$, the electromagnetic field vanishes and the \emph{unique vacuum solution} on the extremal isolated horizon is
\begin{align} \label{th:MetricFunctionNOCHARGES}
  f(\zeta) &= \frac{2}{\pi}\, \frac{(2\pi+\delta_-)(2\pi+\delta_+)(1-\zeta^2)}
  {4\pi(1 + \zeta^2)+ \delta_-(1 + \zeta)^2 +\delta_+(1-\zeta)^2 }.
\end{align}

In addition, \emph{without the deficit angles} ${(\delta_- =0= \delta_+)}$, both poles/axes are regular, and the solution further simplifies to
\begin{align} \label{th:MetricFunctionNODEFICITS}
  f(\zeta) &= 2\, \frac{1-\zeta^2}  {1 + \zeta^2}.
\end{align}
This function corresponds to extremal Kerr black hole \cite{Lewandowski-Pawlowski}, see also the review \cite{KunduriLucietti}.

The geometry of extremal, stationary and axisymmetric isolated horizons (EIHs) is unique in the sense that the induced canonical metric \eqref{CanonicalMetric} must have the metric function in the form of \eqref{th:MetricFunction}. In \cite{Lewandowski-Pawlowski} a similar result was shown for IHs of \emph{regular} spherical topology. The authors initially derived their metric function which would admit conical singularities. However, this additional freedom was removed and the corresponding parameters were not interpreted. In \cite{KunduriLucietti}, the conical singularities are implicitly allowed in the generic formula (80)--(82) when ${c_1 \ne 0}$. The identification is obtained via the relation ${R^2\zeta=x-x_0}$ for a suitably chosen constant $x_0$.

Regular IH necessarily coincides with that of the Kerr-Newman black hole, and forms a 3-parameter class of solutions. By relaxing the regularity condition for the metric function $f(\zeta)$, \emph{we have now obtained a richer 5-parameter class of possible geometries}.
The two additional parameters ${\delta_-}$ and ${\delta_+}$, introduced in (\ref{BoundaryConditions}) as the deficit angles at  ${\zeta = \pm 1}$, are usually interpreted as the cosmic strings (or struts) causing an acceleration of the black hole. It is to be expected intuitively that in the presence of a NUT-like parameter~$l$, these strings/struts would be rotating. Such a possible physical interpretation of the solution \eqref{th:MetricFunction} will be the task of the following section.

We have already pointed out that IHs exhibit specific behaviour, which uniquely defines their structure when they become \emph{extremal}. In the case of the Meissner effect, outer electromagnetic field is repulsed from the horizon. This was proven under weaker assumptions, using the notion of the \emph{almost isolated horizons} \cite{Scholtz2018}. One might expect that imposing the same assumptions the geometry becomes unique, but it is actually not the case. To prove Theorem~\ref{th:MetricIH}, one needs the full time-independence of the isolated horizon expressed by \eqref{eq:LamdbaMuTimeIndependence}. Thus, the Meissner effect is more general and remains valid even when a certain time-dependence is allowed. From the physical point of view, it might be understood as an effect emanating from a different physical feature.

For the sake of completeness, we also derive an explicit result for the $\Psi_2$ projection of the Weyl tensor. The first equation of \eqref{EquationForMetricFunction} combined with \eqref{eq:EquationForPi} yields
\begin{align*}
\Psi_2 \eqNH \pinp \big( 2 a + \pinp - \pinpb\big),
\end{align*}
which is a simple algebraic constraint for $\Psi_2$. When we use the previous results \eqref{ExtremalPiAndPhi}, \eqref{CPhi} and \eqref{eq:SpinCoefficientA}, after some manipulation we arrive at the following  expression in the terms of the integration constant $c_{\pi}$,
\begin{align*}
\Psi_2 \eqNH  - \frac{|Q|^2(c_{\pi}^2 - 1)(\bar{c}_{\pi}^2 - 1)(1 + c_{\pi} \zeta)}{ R^4(\zeta + \bar{c}_{\pi})(\zeta + c_{\pi})^3(1 - c_{\pi} \bar{c}_{\pi})} = f(\zeta) \,\frac{1 + c_{\pi} \zeta}{R^2(1 - \zeta^2)(\zeta + c_{\pi})^2}.
\end{align*}
Direct substitution for the constant $c_{\pi}$ seems to give a somewhat messy formula.

With this choice of the tetrad, it may be also shown that $\Psi_0 \eqNH \Psi_1 \eqNH 0$ on every non-expanding horizon. The projection $\Psi_4$ of the Weyl tensor is a part of the \emph{free data}, and it has to be prescribed on a null hypersurface intersecting the horizon in the spherical-like section $\mathcal{K}$, see  \cite{Krishnan-2012} for details. Finally, the projection $\Psi_3$ is governed by the Ricci identity \eqref{np:RI:deltalambda} and depends on the free data $\lambda$ and $\mu$ as
\begin{align*}
\Psi_3 \eqNH \left(\bar{\eth} + \pinp\right)\mu - \left(\eth + \pinpb\right)\lambda + \Phi_{21} .
\end{align*}

\section{Exact type~D black holes}
\label{sec:exact type D}

A general class of black hole spacetimes of algebraic type D with electromagnetic field (which is not null and is double aligned with the gravitation field) is provided by the Pleba\'{n}ski-Demia\'{n}ski solution~\cite{Plebanski1976}, extending the previous results of Debever~\cite{Debever1971}. It includes, as special cases, the well-known solutions such as the Kerr-Newman black hole, the C-metric or the Taub-NUT solution. Therefore, it appears as the most suitable candidate to compare our main result \eqref{th:MetricFunction} with.

Let us recall the modified form of the Pleba\'{n}ski-Demia\'{n}ski line element first presented in \cite{Griffiths2005} (see also \cite{GriffithsPodolsky2006} or  eq.~(16.5) in the review \cite{Griffiths2009}),
\begin{align}
\dd s^2  = -\frac{1}{(1 - \alpha p r)^2} & \left(  - \frac{\QQ}{r^2 + \omega^2 p^2} \, (\dd \tau - \omega p^2 \dd \sigma)^2 + \frac{r^2 + \omega^2 p^2}{\QQ} \, \dd r^2  \right. \nonumber\\
& \quad + \left. \frac{\orP}{r^2 + \omega^2 p^2} \, (\omega \, \dd \tau + r^2 \dd \sigma)^2 + \frac{r^2 + \omega^2 p^2}{\orP} \, \dd p^2 \right),  \label{PDmetric}
\end{align}
where $\orP (p)$ and $\QQ (r)$ are \emph{polynomials} of the \emph{fourth order}. The metric depends on 9 free parameters, namely $\alpha, \omega, m, n, \epsilon, k, e, g, \Lambda$, from which two can be, in principle, chosen arbitrarily. Direct physical interpretation can be given only to the electric and magnetic charges $e$ and $g$, and the cosmological constant $\Lambda$. The parameter $\alpha$ determines the acceleration of the source, while $\omega$ measures the twist of the (double degenerate) principal null directions.

The null tetrad adapted to these principal null directions of the Weyl tensor reads
\begin{align}
l^a & = \frac{1 - \alpha p r}{\sqrt{2(r^2 + \omega^2 p^2)}} \left( \frac{1}{\sqrt{\QQ}} \left( r^2 \pd^a_{\tau} - \omega \pd^a_{\sigma} \right) - \sqrt{\QQ} \, \pd^a_r \right),\nonumber\\
n^a & = \frac{1 - \alpha p r}{\sqrt{2(r^2 + \omega^2 p^2)}} \left( \frac{1}{\sqrt{\QQ}} \left( r^2 \pd^a_{\tau} - \omega \pd^a_{\sigma} \right) + \sqrt{\QQ} \, \pd^a_r \right), \label{PDNPtetrad}\\
m^a & = \frac{1 - \alpha p r}{\sqrt{2(r^2 + \omega^2 p^2)}} \left( - \frac{1}{\sqrt{\orP}} \left( \omega p^2 \pd^a_{\tau} + \pd^a_{\sigma} \right) + \ii \sqrt{\orP} \,  \pd^a_p \right). \nonumber
\end{align}
In this tetrad, the only non-vanishing curvature scalars are the $\Psi_2$ component of the Weyl tensor, the $\Phi_{11}$ component of the Ricci tensor, and the Ricci scalar $R$ which has the value ${R=4\Lambda}$, for more details see \cite{Griffiths2009}. Explicit expressions for these quantities indicate the presence of curvature singularity at ${r=0=\omega p}$. The function $\orP$ must be positive, and ${\orP(p)=0}$ identifies the poles (axes of symmetry). The horizons are determined by a condition ${\QQ(r_H) = 0}$.

In order to get a line element that explicitly represents the family of black holes, identifies the physical meaning of the free parameters, and reduces to the well-known solutions by setting these parameters to zero,  the authors of \cite{Griffiths2005} employed a coordinate transformation with two additional parameters $a$ and $l$, namely
\begin{align*}
\tau = t - \frac{(l+a)^2}{a}\, \fii, \qquad
p = \frac{l}{\omega} + \frac{a}{\omega}\,\varsig, \qquad
\sigma=- \frac{\omega}{a}\, \fii.
\end{align*}
The metric \eqref{PDmetric} is thus transformed into (c.f. \cite{Griffiths2005}, \cite{GriffithsPodolsky2006} or  eq.~(16.12) in \cite{Griffiths2009})
\begin{align}
\dd s^2 = -\frac{1}{\Omega^2} & \bigg( -\frac{\QQ}{\rho^2}\left(\dd t - \left[ a (1-\varsig^2) + 2l\,(1-\varsig) \right] \dd \fii \right)^2 + \frac{\rho^2}{\QQ}\,\dd r^2  \nonumber\\
& \quad  + \,\frac{\rho^2}{\tilde{P}} \, \dd \varsig^2 + \frac{\tilde{P}}{\rho^2} \left[ a\, \dd t - \big(r^2 + (a+l)^2\big)\, \dd \fii \right]^2 \bigg), \label{PDmetricGP}
\end{align}
where
\begin{align*}
& \Omega = 1 - \alpha\, \Big(\,\frac{l}{\omega}+\frac{a}{\omega}\, \varsig\Big)\, r, \qquad
  \rho^2 = r^2 + (l + a\, \varsig)^2, \\
& \tilde{P} (\varsig) = a_0 + a_1 \,\varsig + a_2 \,\varsig^2 + a_3 \,\varsig^3 + a_4 \,\varsig^4,\\
& \QQ (r) = \, b_0 + b_1 \,r + b_2 \,r^2 + b_3 \,r^3 + b_4 \,r^4.
\end{align*}
The mutually related constants ${a_i, b_i}$ are specific combinations of the initial  Pleba\'{n}ski-Demia\'{n}ski parameters (for their explicit form see \cite{Griffiths2005}, \cite{GriffithsPodolsky2006} or eqs.~(16.12)--(16.13) in \cite{Griffiths2009}).
The transformed NP tetrad (\ref{PDNPtetrad}) now reads\footnote{Notice that the vector $\vec{m}$ involves the original polynomial $\orP$, not $\tilde{P}$.}
\begin{align}
l^a &= \frac{\Omega}{\sqrt{2 \QQ}\, \rho}\Big( \big[r^2+(a+l)^2\big] \pd^a_t + a\, \pd^a_{\fii}
  - \QQ\, \pd^a_r
  \Big),\nonumber \\
n^a &= \frac{\Omega}{\sqrt{2 \QQ}\, \rho}\Big( \big[r^2+(a+l)^2\big] \pd^a_t + a\, \pd^a_{\fii}
  + \QQ\, \pd^a_r
  \Big), \label{eq:Initialn}\\
m^a &= \frac{\Omega\,a}{\sqrt{2 \orP}\, \rho\, \omega} \Big( (1-\varsig)\big[\,a(1+\varsig) + 2l\,\big] \pd^a_t + \pd^a_{\fii} + \ii \, \orP\, \frac{\omega^2}{a^2} \,\pd^a_{\varsig} \Big). \nonumber
\end{align}

The classic black hole solutions are identified in the large class of spacetimes (\ref{PDmetricGP}) when the polynomial $\tilde{P}$ has the particular factorized form
\begin{align}
\tilde{P}(\varsig) = (1 - \varsig^2)(1 - a_3\, \varsig - a_4\,\varsig^2) \label{eq:PolynomialPtilde}
\end{align}
with \emph{two distinct roots} (poles) at ${\varsig = \pm 1}$.  Then the coordinate $\fii$ might be recognized as a \emph{periodic coordinate with respect to the axes located at} ${\varsig = \pm 1}$.

With this choice, it is also natural to consider
\begin{align}
\varsig = \cos\theta, \qquad \theta\in[0,\pi]. \label{eq:VarsigTheta}
\end{align}
Introducing ${P(\varsig) \equiv \tilde{P}(\varsig) / (1 - \varsig^2)}$ and \emph{assuming}~${\Lambda=0},$\footnote{Generalization to an any value of the cosmological constant $\Lambda$ was presented in  \cite{PodolskyGriffiths2006} and \cite{GriffithsPodolsky2006}.}  the metric (\ref{PDmetricGP}) with (\ref{eq:PolynomialPtilde}) then takes the explicit form (see the line element (14) in~\cite{Griffiths2005})
\begin{align}
\dd s^2 = -\frac{1}{\Omega^2} &
  \left(-\frac{\QQ}{\rho^2}\left[\dd t- \left(a\sin^2\theta +4l\sin^2\!{\textstyle\frac{1}{2}\theta} \right)\dd\fii \right]^2 + \frac{\rho^2}{\QQ}\,\dd r^2 \right. \nonumber\\
& \quad \left. + \,\frac{\rho^2}{P}\,\dd\theta^2
  + \frac{P}{\rho^2}\,\sin^2\theta\, \big[ a\dd t -\big(r^2+(a+l)^2\big)\,\dd\fii \big]^2
 \right), \label{newmetricGP2005}
\end{align}
where
\begin{align}
& \Omega = 1 - \alpha\, \Big(\,\frac{l}{\omega}+\frac{a}{\omega}\cos\theta \Big)\, r, \qquad
  \rho^2 = r^2 + (l + a \cos\theta)^2, \nonumber\\
& P(\theta) = 1-a_3\cos\theta-a_4\cos^2\theta, \nonumber\\
&
\QQ (r) = \Big[(\omega^2k+e^2+g^2)\Big(1+2\alpha\,\frac{l}{\omega}\,r\Big)
  -2m\,r +\frac{\omega^2k}{a^2-l^2}\,r^2\Big]
  \Big[1+\alpha\,\frac{a-l}{\omega}\,r\Big] \Big[1-\alpha\,\frac{a+l}{\omega}\,r\Big],
  \label{factorizedQ}
\end{align}
with
\begin{align}
a_3 &= 2\alpha \,\frac{a}{\omega}\,m -4\alpha^2\, \frac{a\,l}{\omega^2}\,  (\omega^2k+e^2+g^2),  \nonumber\\
a_4 &= -\alpha^2\,\frac{a^2}{\omega^2}\,(\omega^2k+e^2+g^2), \label{a34}
\end{align}
and $\omega^2 k$ given by
\begin{equation}
  \frac{\omega^2 k}{a^2-l^2} =
  \frac{{\displaystyle 1 +2\alpha\,\frac{l}{\omega}\,m -3\alpha^2\frac{l^2}{\omega^2}\,(e^2+g^2)}}
  {{\displaystyle 1 + 3\alpha^2\,\frac{l^2}{\omega^2}\,(a^2-l^2)}},
  \label{k}
\end{equation}
so that
\begin{equation}
  \omega^2 k+e^2+g^2 =
  \frac{{\displaystyle (a^2-l^2 +e^2+g^2) +2\alpha\,\frac{l}{\omega}(a^2-l^2)\,m }}
  {{\displaystyle 1 + 3\alpha^2\,\frac{l^2}{\omega^2}\,(a^2-l^2)}}.
  \label{keg}
\end{equation}

The metric (\ref{newmetricGP2005}) explicitly depends on \emph{six usual physical parameters} $m$, $a$, $l$, $\alpha$, $e$, $g$ which characterize mass, Kerr-like rotation, NUT parameter, acceleration, electric and magnetic charges of the black hole, respectively.

The additional twist parameter $\omega$ is free in the sense that the remaining coordinate freedom can be used to set $\omega$ to any convenient value if at least one of the parameters $a$ or $l$ are non-zero (otherwise ${\omega\equiv0}$), see the discussion in \cite{Griffiths2005, GriffithsPodolsky2006}. In particular, it is natural to set ${\omega = a}$ when  ${l = 0}$. An interesting gauge choice of $\omega$ was recently suggested in \cite{Vratny2018}, namely
\begin{align}
\omega \equiv \frac{a^2 + l^2}{a}, \label{eq:ChoiceOfOmega}
\end{align}
so that
\begin{align}
\frac{a}{\omega} = \frac{a^2}{a^2 + l^2}, \qquad \frac{l}{\omega} = \frac{a\,l}{a^2 + l^2} . \label{eq:ChoiceOfOmega-al}
\end{align}
With this choice, the general metric \eqref{newmetricGP2005} reduces directly to the familiar forms of either the Kerr--Newman, the Taub-NUT solution or the $C$-metric in appropriate cases, without the need for further transformations, simply by setting the corresponding parameters to zero.

The metric (\ref{newmetricGP2005}) is also convenient for identifying the \emph{horizons}. They are located at such values of the radial coordinate ${r = r_H}$ which satisfy the condition
\begin{align}
\QQ (r_H) = 0. \label{eq:Horizons}
\end{align}
Since the function $\QQ(r)$ given by (\ref{factorizedQ}) is factorized, the corresponding roots are immediately seen. There are \emph{two acceleration horizons} located at
\begin{equation}
r_{{\rm a}+} = \frac{a^2 + l^2}{\alpha\, a\, (a+l)}, \qquad r_{{\rm a}-} = -\frac{a^2 + l^2}{\alpha\, a\, (a-l)},
  \label{accel-horizons}
\end{equation}
and (in general) \emph{two black hole horizons} located  at the roots of the first square bracket in (\ref{factorizedQ}),
\begin{equation}
\frac{\omega^2k}{a^2-l^2}\,r_H^2 - 2\Big[ m - \alpha\,\frac{l}{\omega}(\omega^2k+e^2+g^2)\Big]\,r_H
+(\omega^2k+e^2+g^2) = 0.
  \label{BHrootsofQ}
\end{equation}
The degenerate case when the \emph{two horizons coincide}, identifying the \emph{extremal} black hole horizon~$r_H$, corresponds to the vanishing discriminant. This explicitly reads
\begin{align}
r_H &=\Big[ m - \frac{\alpha\,a\,l}{a^2 + l^2}(\omega^2k+e^2+g^2)\Big]{\frac{a^2-l^2}{\omega^2k}}  \nonumber\\
   &\quad\Leftrightarrow\quad
 \Big[ m - \frac{\alpha\,a\,l}{a^2 + l^2}(\omega^2k+e^2+g^2)\Big]^2=\frac{\omega^2k}{a^2-l^2}(\omega^2k+e^2+g^2).
  \label{extremerootofQ}
\end{align}

In the \emph{absence of acceleration}~$\alpha$ or for \emph{vanishing rotation}~$a$ or for \emph{vanishing NUT parameter}~$l$ (i.e., when ${\alpha\,a\,l=0}$), the quadratic equation \eqref{BHrootsofQ} simplifies considerably to ${r_H^2 - 2 m \,r_H +(a^2-l^2 +e^2+g^2) = 0}$, the  roots are ${\,r_\pm\equiv m \pm \sqrt{m^2-a^2+l^2-e^2-g^2}\,}$, and the extremal horizon is at
\begin{equation}
r_H=m  \qquad\Leftrightarrow\qquad m^2+l^2=a^2+e^2+g^2.
  \label{extremeroot-alpha=0nebol=0}
\end{equation}

\vspace{2mm}

\subsection{Calculation of the horizon geometry}

The first step towards the derivation of the isolated horizon structure in the class of black hole spacetimes (\ref{newmetricGP2005}) is to find a horizon generator $l^a_H$ and introduce the advanced time coordinate $v$ in such a way that ${l^a_H \eqNH \pd_v^a}$, in agreement with \eqref{eq:AdaptedCoordinates}.

Moreover, the theory of isolated horizons requires that both~$v$ and the spatial coordinates~${x^I}$, ${I \in \{1,2\}}$, are parallelly propagated along the vector $n^a$,
\begin{align}
\Delta v = \Delta x^I = 0, \label{eq:ConditionsOnCoordinates}
\end{align}
c.f.~\eqref{eq:PropagationOfCoordinates}. From the expression \eqref{eq:Initialn} we see that the coordinate $\varsig$ already satisfies this requirement but $\fii$ does not. Hence, we perform the transformation of coordinates ${v = v(t, r)}$, ${\tilde{\phi} = \tilde{\phi}(\fii, r)}$. The transformation ensuring the conditions \eqref{eq:ConditionsOnCoordinates} may be chosen in the following way\footnote{In general, one gets from \eqref{eq:ConditionsOnCoordinates} the relation ${\pd_r v = -n^t/n^r \, \pd_t v}$, and we choose ${\pd_t v = 1}$. The coordinate $\tilde{\phi}$ is fixed analogously.}
\begin{align*}
\dd v = \dd t - \frac{r^2 + (a+l)^2}{\QQ(r)} \,\dd r, \qquad \dd \tilde{\phi} = \dd \fii - \frac{a}{\QQ(r)} \,\dd r.
\end{align*}
In the new coordinates ${\{ v, r, \varsig, \tilde{\phi} \}}$, the NP tetrad (\ref{eq:Initialn})  reads
\begin{align}
l^a &= \frac{\Omega}{\sqrt{2\QQ}\, \rho} \Big( 2\big[r^2+(a+l)^2\big]\pd^a_v
+ 2a\, \pd^a_{\tilde{\phi}} - \QQ \,\pd^a_r  \Big),\nonumber\\
n^a &= \frac{\Omega\sqrt{\QQ} }{\sqrt{2}\, \rho} \,\pd^a_r, \label{eq:nlmTRANSFORMED}\\
m^a &= \frac{\Omega\,a}{\sqrt{2 \orP}\, \rho \, \omega} \Big( (1 - \varsig)\big[\,a(1+\varsig) + 2l\,\big] \,\pd^a_v + \pd^a_{\tilde{\phi}} + \ii \,\orP\, \frac{ \omega^2}{a^2} \,\pd^a_{\varsig} \Big),\nonumber
\end{align}
and the line element \eqref{PDmetricGP} becomes
\begin{align} \label{eq:FullMetric}
\dd s^2 = &\frac{1}{\Omega^2\rho^2 }\bigg( (\QQ - a^2 \tilde{P})\dd v^2 + 2\rho^2 \dd v \, \dd r
+ 2(\varsig-1) \rho^2 \big[\,a(1+\varsig) + 2l\,\big]\dd r \, \dd \tilde{\phi}
 \nonumber\\
&\qquad + 2 \Big[ \QQ(\varsig - 1)\big[\,a(1+\varsig) + 2l\,\big] + a \tilde{P} \big[r^2+(a+l)^2\big] \Big] \dd v \, \dd \tilde{\phi}  - \frac{\rho^4}{\tilde{P}} \dd \varsig^2 \\
&\qquad + \left[ \QQ(\varsig - 1)^2\big[\,a(1+\varsig) + 2l\,\big]^2 - \tilde{P} \big[r^2+(a+l)^2\big]^2 \right] \dd \tilde{\phi}^2 \bigg), \nonumber
\end{align}
where $\tilde{P}(\varsigma)$ takes the form (\ref{eq:PolynomialPtilde}) with (\ref{a34}). The horizons are located at ${r = r_H}$ such that ${\QQ (r_H) = 0}$, see (\ref{eq:Horizons}). The vector $l^a$ in the transformed NP tetrad (\ref{eq:nlmTRANSFORMED}) thus diverges on any horizon. To get rid of this divergence, we define a \emph{rescaled} normal to the horizon as
\begin{align*}
l^a_H \equiv \pd^a_v + \frac{a}{r^2+(a+l)^2} \,\pd^a_{\tilde{\phi}} - \frac{\QQ}{ 2\big[r^2+(a+l)^2\big]}\, \pd^a_r  \,\eqNH \,\pd^a_v + \frac{a}{r_H^2+(a + l)^2}\, \pd^a_{\tilde{\phi}}\,.
\end{align*}
Using~\eqref{eq:FullMetric}, one can easily check that $l^a_H$ is indeed null on the horizon. The corresponding null vector $n^a_H$ is obtained simply by the scaling\footnote{The vector $m^a$ is fixed, so if ${\tilde{l^a} = c\, l^a}$ then ${\tilde{n^a} = c^{-1} n^a}$ to keep the normalization.}
\begin{align*}
n^a_H = \frac{\Omega^2}{\rho^2} \big[r^2+(a+l)^2\big]\,\pd^a_r.
\end{align*}
However, the vector $l^a_H$ does not yet have the required form. To achieve ${l^a_H \eqNH \pd^a_v}$, we introduce another transformation of coordinates $\bar{\phi} = \bar{\phi}(v, \tilde{\phi})$, which reads
\begin{align*}
\dd \bar{\phi} = \dd \tilde{\phi} - \frac{a}{r_H^2 + (a + l)^2}\, \dd v.
\end{align*}
Notice that the value of the radial coordinate is here fixed at $r_H$. The normal $l^a_H$ is thus transformed into ${l^a_H \eqNH \pd^a_v}$, as required, while $n^a_H$ remains the same, and $m^a$ is transformed into $m^a_H$ (which is not necessary to explicitly write here).

Let us also observe that the line element on a section of the horizon ${r=r_H}$ for ${v=\hbox{const.}}$ reads
\begin{align} \label{eq:HorizonMetric}
\dd s_H^2 = - \frac{\rho^2}{\Omega^2 \tilde{P}} \, \dd \varsig^2
   - \big[r_H^2+(a+l)^2\big]^2\frac{\tilde{P}}{\Omega^2\rho^2 } \, \dd \bar{\phi}^2.
\end{align}
The range of the angular coordinate $\bar{\phi}$ is not obvious from our construction. We will thus generally assume that $\bar{\phi} \in [0,2 \pi C)$, where we have introduced a ``conicity'' parameter~$C$. Its value will be determined later.

\subsection{Evaluation of the surface gravity}

The surface gravity $\surg$ is defined as the ``acceleration'' of the null normal of the horizon,
\begin{align}
l^a_H \nabla_a (l_H)_b \eqNH \surg (l_H)_b.
\label{eq:surfacegravity}
\end{align}
The covariant form of the normal ${l^a_H \eqNH \pd^a_v}$ on the horizon reads
\begin{align*}
(l_H)_a \eqNH \frac{\rho^2}{\Omega^2 [r_H^2+(a+l)^2\big]}\, \dd r_a.
\end{align*}
Substitution into the equation \eqref{eq:surfacegravity} yields
\begin{align*}
& \surg (l_H)_r \delta_{r b } \eqNH l_H^v \nabla_v (l_H)_b \eqNH \pd_v (l_H)_b - \Gamma^c_{\, v b} (l_H)_c \eqNH - \Gamma^r_{\, v b} (l_H)_r \\
& \Rightarrow \quad \surg \eqNH - \Gamma^r_{\, v r}, \quad \Gamma^r_{\, v b} \eqNH 0, \  \forall b \neq r.
\end{align*}
For the metric~\eqref{eq:FullMetric}, the surface gravity turns out to be
\begin{align*}
\surg = - \frac{\QQ'(r_H)}{2\big[r_H^2+(a+l)^2\big]},
\end{align*}
where the prime denotes the derivative with respect to the argument $r$. The second condition ${\Gamma^r_{\, v b} \eqNH 0, \, \forall b \neq r}$ is also satisfied.

\emph{Extremal horizons} are characterized by \emph{vanishing of their surface gravity}. Hence, for the extremal case we have the condition
\begin{align}
\QQ' (r_H) = 0. \label{eq:ConditionOfExtremality}
\end{align}
In view of the factorized form (\ref{factorizedQ}) of the metric function $\QQ(r)$, it can be seen that the acceleration horizons at ${r_{{\rm a}+}}$ and ${r_{{\rm a}-}}$ (which are present when ${\alpha \ne 0}$) can not be extremal because their surface gravity is non-zero.\footnote{The hypothetical case ${r_{{\rm a}+}=r_{{\rm a}-}}$ requires ${a=0}$, ${\alpha\ne0}$, ${l\ne0}$, which is the accelerating NUT black hole. As shown recently in \cite{PodolskyVratny2020}, such spacetimes are algebraically general and thus do not belong to the investigated family of type~D solutions. Even in such a case, there are no extreme black hole horizons.}
The only possibility to obtain an extremal horizon is \emph{when the two black-hole horizons coincide}.
It can easily be shown that the extremality condition \eqref{eq:ConditionOfExtremality} is fully consistent with the explicit solution~\eqref{extremerootofQ} for the position $r_H$ of the two coinciding black-hole horizons.

\subsection{Expansion and twist of the null normal}

The null normal to an isolated horizon is required to be expansion-free and twist-free. Both these properties are encoded in the spin coefficient $\rho$ defined as (see, e.g., \cite{Matejov-2018})
\begin{align*}
\rho \equiv m^a\, \conj{\delta} l_a \equiv m^a \,\conj{m}^b\, \nabla_b \,l_a.
\end{align*}
Namely, the expansion $\expl$ is its real part, while the twist $\upomega$ is its imaginary part,
\begin{align*}
\expl = 2\, \Re{\rho}, \quad \upomega = \sqrt{2}\, \Im{\rho}.
\end{align*}
Using the definition of $\rho$ above, we calculate the expansion and twist of the vector $l^a_H$ on the horizon, yielding
\begin{align*}
\rho \eqNH 0.
\end{align*}
The vector field $l^a_H$ is thus indeed non-expanding and has zero twist on the horizon (so it has a Kundt-like property).

\subsection{Geometry of the horizon sections}

The line element on a spherical-like section of the horizon $\mathcal{K}$  for ${r=r_H}$ and ${v=\hbox{const.}}$ is given by \eqref{eq:HorizonMetric}. In order to compare this metric with \eqref{CanonicalMetric}, we first rescale the angular coordinate $\bar{\phi}$ as
\begin{align*}
\bar{\phi} = C \phi,
\end{align*}
so it acquires its values in the required range ${\phi \in [0,2\pi)}$. The line element is then
\begin{align} \label{eq:HorizonMetricII}
\dd s_H^2 = - \frac{\rho^2}{\Omega^2 \tilde{P}} \, \dd \varsig^2
   - C^2 \big[r_H^2+(a+l)^2\big]^2\frac{\tilde{P}}{\Omega^2\rho^2 } \, \dd \phi^2,
\end{align}
where~$C$ is the additional free \emph{conicity parameter}. However, it still does not have the canonical form \eqref{CanonicalMetric}. The next step towards the desired form is to calculate the horizon area~$A$, which is required for determining the radius $R$ via ${A=4\pi R^2}$.

The invariant volume element corresponding to~\eqref{eq:HorizonMetricII} reads
\begin{align*}
\text{vol}(\mathcal{K}) =  C\, \frac{r_H^2+(a+l)^2}{\Omega^2} \, \dd \varsig \, \dd \phi,
\end{align*}
and thus the area is
\begin{align}\label{eq:HorizonArea}
A \equiv \oint_{\mathcal{K}} \text{vol}(\mathcal{K})
  &= \int_{-1}^1 \int_0^{ 2 \pi}  C\, \frac{r_H^2+(a+l)^2}{\Omega^2} \, \dd \varsig \, \dd \phi \nonumber\\
  &= \frac{ 4\pi C\, \big[r_H^2+(a+l)^2\big]}
{\Big[1-\alpha\, r_H{\displaystyle \Big(\frac{a}{\omega} + \frac{l}{\omega} \Big)}\Big]
 \Big[1+\alpha\, r_H{\displaystyle \Big(\frac{a}{\omega} - \frac{l}{\omega} \Big)}\Big]}.
\end{align}

By comparing \eqref{eq:HorizonMetricII} with \eqref{CanonicalMetric}, we infer that the azimuthal coordinates on the horizon are related by a coordinate transformation ${\zeta= \zeta(\varsig)}$ such that
\begin{align}
\dd \zeta = \frac{4 \pi C}{A} \frac{ r_H^2+(a + l)^2}
{\Omega^2} \, \dd \varsig    \qquad\hbox{where}\qquad
\Omega(\varsig) = 1 - \alpha\,r_H {\displaystyle \Big(\,\frac{l}{\omega} + \frac{a}{\omega}\, \varsig \Big) } . \label{eq:TransformationOfZeta}
\end{align}

Moreover, the azimuthal coordinate $\zeta$ adapted to an axially symmetric isolated horizon has to satisfy the condition
\begin{align*}
\oint_{{\mathcal{K}} } \zeta \, \text{vol}(\mathcal{K}) = 0,
\end{align*}
which effectively fixes the integration constant in \eqref{eq:TransformationOfZeta}. Recall that $\zeta$ is constructed as a solution to a certain differential equation, see \cite{Ashtekar2004-multipolemoments}. To obtain an unambiguous solution, one has to employ the condition above. The range of $\zeta$ is then fixed by definition of $R^2$ to $\zeta \in [-1,1]$, c.f. equation \eqref{CanonicalMetric} and discussion therein.

After integration we obtain
\begin{align}
\zeta(\varsig) =
\frac{\varsig - \alpha\, r_H {\displaystyle \Big(\,\frac{a}{\omega} + \frac{l}{\omega}\, \varsig \Big)}}
     {1 - \alpha\,r_H {\displaystyle \Big(\,\frac{a}{\omega}\, \varsig + \frac{l}{\omega} \Big) }}
\qquad \Rightarrow \qquad
\varsig(\zeta) = \frac{ \zeta + \alpha\, r_H {\displaystyle \Big(\,\frac{a}{\omega} - \frac{l}{\omega}\, \zeta \Big)}}{1 + \alpha\,r_H {\displaystyle \Big(\,\frac{a}{\omega}\, \zeta - \frac{l}{\omega} \Big) }}. \label{eq:ZetaAndVarsigma}
\end{align}
Both coordinates are in the required range ${\varsig, \zeta \in[-1,1]}$. Moreover, ${\zeta(\varsig=\pm 1) = \pm 1}$.

To have a well-defined coordinate, $\zeta$ is required to be an increasing function of $\varsig$, see \cite{Ashtekar2004-multipolemoments}. Therefore, the derivative has to be positive,
\begin{align} \label{eq:ConditionOnParameters}
\zeta'(\varsig) > 0 \qquad \Rightarrow \qquad
 1 - \alpha\, r_H \Big(\,\frac{a}{\omega} + \frac{l}{\omega} \Big) > 0,
\end{align}
which puts a restriction on the black hole parameters. For a given Kerr parameter~$a$ and the NUT parameter~$l$, the acceleration $\alpha$ \emph{can not be too large}. In fact, since the acceleration horizon $r_{{\rm a}+}$ is defined by the condition  ${\alpha\,r_{{\rm a}+}(\frac{a}{\omega} + \frac{l}{\omega})=1}$, see~\eqref{factorizedQ}, the constraint \eqref{eq:ConditionOnParameters} can be rewritten simply as ${r_H<r_{{\rm a}+}}$, which is naturally always satisfied.

Let us also remark that the acceleration horizons \eqref{accel-horizons} are related as
\begin{align*}
 r_{a+} = \frac{l-a}{l+a}\, r_{a-}.
\end{align*}
In the case when ${a=0}$ these horizons coincide, while in the ${l=0}$ case they differ by a sign,  ${r_{a+} = - r_{a-}}$. Depending on the particular values of  $a$ and $l$, their signs and the sign of ${r_{a-}}$, mutual positions of the acceleration horizons may be $r_{a+} \gtreqless r_{a-}$.

After the transformation of coordinates \eqref{eq:ZetaAndVarsigma}, the line element \eqref{eq:HorizonMetricII} is recast into the canonical form \eqref{CanonicalMetric}, and we thus arrive at our another \emph{key result}:

\begin{theorem} \label{th:MetricD}
The specific metric function $f_{\text{D}}(\zeta)$, which decribes the geometry of the horizon in a complete family of type~D black holes \eqref{newmetricGP2005}, is given by
\begin{align}
f_{\text{D}}(\zeta) &= \frac{4\pi C^2}{A} \big[r_H^2+(a + l)^2\big]^2
    \frac{\tilde{P}(\zeta)}{\Omega^2(\zeta)\,\rho^2(\zeta)},
\label{eq:MetricFunction}
\end{align}
where the functions $\tilde{P}, \Omega, \rho$, introduced in \eqref{eq:PolynomialPtilde}, \eqref{PDmetricGP}, have to be regarded as functions of the new variable $\zeta$ via \eqref{eq:ZetaAndVarsigma}, for example  ${\rho(\zeta) \equiv \rho\big(\varsig(\zeta)\big)}$. From equation \eqref{eq:PolynomialPtilde} it follows that ${f_{\text{D}}(\pm 1) = 0}$.
\end{theorem}

A direct substitution and evaluation leads, in general, to a complicated expression, which will be discussed in the following sections.

\subsection{Deficit angles and conicity}
\label{sec:deficit angles conicity}

Recall that for the metric in the form \eqref{CanonicalMetric}, one obtains the limit
\begin{align}
\lim_{\zeta \rightarrow \pm 1} \frac{O(\zeta)}{\varrho(\zeta)} = \mp \pi\, f'(\pm 1), \label{eq:ConicityDefinition}
\end{align}
where $O(\zeta)$ is the \emph{circumference} of a curve defined to have the same value of the coordinate~$\zeta$, while $\varrho(\zeta)$ is the \emph{distance} from the north or south pole, respectively, to a specific point on this curve.\footnote{Since we deal with axially symmetric spacetimes, there is no difference in the proper length of ``radius'' $\varrho$ according to different choice of a point on the curve. But we assume that the difference between $\rho$ and the geodesic ``radius'', which should have been used, is of a higher order and disappears in the limit.} Regular axes are clearly defined by the condition $f'(\pm 1) = \mp 2$. If this is violated, conical singularities are present, see expression~\eqref{BoundaryConditions}.

Straightforward calculation using \eqref{eq:ConicityDefinition} for the metric function \eqref{eq:MetricFunction} for a generic type~D black hole reveals non-zero deficit angles around both poles. With the notation introduced in~\eqref{BoundaryConditions}, these deficit angles at the two poles on the horizon $r_H$ are explicitly given as
\begin{align}
\begin{split}
\delta_+ &= 2\pi \Big(C\,(1 - a_3 - a_4) - 1 \Big),\\
\delta_- &= 2\pi \Big(C\,(1 + a_3 - a_4)\,\frac{r_H^2+(a+l)^2}{r_H^2+(a-l)^2} - 1 \Big),
\label{eq:DeficitAngles}
\end{split}
\end{align}
where the constants ${a_3, a_4}$ are given by \eqref{a34} and $C$ is the free conicity parameter.

Moreover, it is now also clear that for a unique choice of the conicity parameter $C$ we can always achieve a \emph{vanishing deficit angle at one of the poles}. For example,
\begin{align}
\delta_+ =0  \qquad \Leftrightarrow \qquad  C = (1 - a_3 - a_4)^{-1}, \label{eq:DeficitAngles+=0}
\end{align}
in which case (generically) ${\delta_- \ne 0}$.

It is also possible to achieve ${\delta_- = 0}$ but, in this case, one has to be more careful. When ${l \ne 0}$, the metric (\ref{PDmetricGP}) is \emph{not} regular at~${\varsigma=-1}$ (since ${g_{\fii\fii}\ne 0}$). It is necessary to regularize this axis/pole, and it is first achieved be changing the time coordinate appropriately. In particular, by performing the transformation ${t \to \bar t}$ such that
\begin{align}
t = \bar t + 4l\, \varphi, \label{eq:tbart}
\end{align}
the metric (\ref{PDmetricGP}) becomes
\begin{align}
\dd s^2 = -\frac{1}{\Omega^2} & \bigg( -\frac{\QQ}{\rho^2}\left(\dd \bar t - \left[ a (1-\varsig^2) - 2l\,(1+\varsig) \right] \dd \fii \right)^2 + \frac{\rho^2}{\QQ}\,\dd r^2 . \nonumber\\
& \quad  + \,\frac{\rho^2}{\tilde{P}} \, \dd \varsig^2 + \frac{\tilde{P}}{\rho^2}
\left[ a\, \dd \bar t - \big(r^2 + (a-l)^2\big)\, \dd \fii \right]^2 \bigg), \label{PDmetricGPbar}
\end{align}
which admits a regular axis at ${\varsigma=-1}$.

Repeating now the same arguments and calculations as before, we obtain expressions analogous  to (\ref{eq:MetricFunction}) and (\ref{eq:TransformationOfZeta}), namely
\begin{align}
\bar f_{\text{D}}(\zeta) &= \frac{4\pi C^2}{A} \big[r_H^2+(a - l)^2\big]^2
    \frac{\tilde{P}(\zeta)}{\Omega^2(\zeta)\,\rho^2(\zeta)},
\label{eq:MetricFunctionbar}
\end{align}
and
\begin{align}
\dd \zeta = \frac{4 \pi C}{A} \frac{ r_H^2+(a - l)^2}
{\Omega^2} \, \dd \varsig   . \label{eq:TransformationOfZetabar}
\end{align}
The corresponding deficit angles at the two poles on the horizon $r_H$ are now given by
\begin{align}
\begin{split}
\bar\delta_+ &= 2\pi \Big(C\,(1 - a_3 - a_4)\,\frac{r_H^2+(a-l)^2}{r_H^2+(a+l)^2} - 1 \Big), \\
\bar\delta_- &= 2\pi \Big(C\,(1 + a_3 - a_4) - 1 \Big),
\label{eq:DeficitAnglesBar}
\end{split}
\end{align}
so that
\begin{align}
\bar\delta_- =0  \qquad \Leftrightarrow\qquad  C = (1 + a_3 - a_4)^{-1}, \label{eq:DeficitAnglesbar-=0}
\end{align}
in which case (generically) ${\bar\delta_+ \ne 0}$.

These results fully agree with those obtained in~\cite{Griffiths2005, GriffithsPodolsky2006} for the regularity of the poles/axes at ${\theta=0}$ and ${\theta=\pi}$, respectively, using the metric form \eqref{newmetricGP2005}.

\section{Comparison of the results on EIH with the family of type~D black holes}
\label{sec:comparison}

For a general axially symmetric extremal isolated horizon (EIH) we have derived in Sec.~\ref{sec:extremal isolated horizons} the explicit formula \eqref{th:MetricFunction}
for the metric function $f(\zeta)$ in the canonical coordinates $(\zeta, \phi)$.  We will denote it here by $f_{\text{EIH}}(\zeta)$. It reads
\begin{align} \label{eq:MetricFunctionIH}
  f_{\text{EIH}}(\zeta) &= \frac{2}{\pi}\,
  \frac{(2\pi+\delta_-)(2\pi+\delta_+)(1-\zeta^2)}
  {4\pi(1 + \zeta^2)+ \delta_-(1 + \zeta)^2 +\delta_+(1-\zeta)^2 + q^2 (1-\zeta^2)},
\end{align}
where $\delta_{\pm}$ are the \emph{deficit angles on the horizon poles} at ${\zeta=\pm1}$. Recall that, due to \eqref{eq:ZetaAndVarsigma}, these correspond to ${\varsigma=\pm1}$ in the Pleba\'{n}ski-Demia\'{n}ski  metric \eqref{PDmetricGP}, or equivalently to the poles at ${\theta=0}$ and $\pi$, see \eqref{eq:VarsigTheta}. Here we have also introduced a shorthand notation for the \emph{dimensionless combination of charges}
\begin{align} \label{eq:Definiceq^2}
  q^2 \equiv \frac{(4\pi)^2}{A}\,(Q_E^2 + Q_M^2),
\end{align}
where $Q_E$ and $Q_M$ are the electric and magnetic charges, respectively. We also prefer to use the area~${A=4\pi R^2}$ instead of the ``Euclidean'' radius~$R$.

Our aim now is to compare this general result with the family of extremal black holes, which are exact spacetimes of type D, namely with the metric function $f_{\text{D}}(\zeta)$ given in \eqref{eq:MetricFunction}. We wish to clarify the relation between the geometrical (or topological) features and the physical parameters.

The metric function $f_{\text{EIH}}$ given by \eqref{eq:MetricFunctionIH} depends on 5 independent parameters, namely $(A, Q_E, Q_M, \delta_-, \delta_+)$, while the family of algebraic type D black holes \eqref{newmetricGP2005} is characterized by 6 physical parameters $(m, e, g, a, l, \alpha)$. However, one of these parameters is fixed by the condition of extremality of the horizon. Thus, both generic solutions contain \emph{the same number of independent parameters}. At least at first glance, they should match.

First, the deficit angles $\delta_+$ and $\delta_-$, being defined geometrically, \emph{are the same} both in the metric function \eqref{eq:MetricFunctionIH} and  in the type~D metric \eqref{newmetricGP2005}, explicitly determined by the physical parameters via \eqref{eq:DeficitAngles} or \eqref{eq:DeficitAnglesBar}. Recall also that there is an additional free \emph{conicity parameter}~$C$, which can be used to prescribe either the deficit angle $\delta_+$ or the deficit angle $\delta_-$ to \emph{any value}, in particular ${\delta_+=0}$ or ${\delta_-=0}$, see expressions \eqref{eq:DeficitAngles+=0} or \eqref{eq:DeficitAnglesbar-=0}.

The mass parameter $m$ is related to the value of the radial coordinate $r_H$ which defines the horizon, via the condition~\eqref{eq:Horizons}, leading to \eqref{extremerootofQ} and \eqref{extremeroot-alpha=0nebol=0}. On the other hand, the area of the horizon $A$ depends on $r_H$, see \eqref{eq:HorizonArea}, so the mass parameter $m$ is related to the area~$A$. Moreover, we naturally assume that both the electric and magnetic charges $Q_E, Q_M$ are related to their counterparts $e, g$ (via the relations yet to be determined).

Generally, it is natural to make the following assumptions and restrictions on these parameters, namely:
\begin{enumerate}

\item The horizon area $A$ and the mass parameter $m$ are non-negative (${A \geq 0 }$, ${m \geq 0}$).

\item The charges $Q_E, Q_M$ are determined by the Gauss law.

\item The deficit angles are in the range ${\delta_{\pm} \in (-2\pi, \infty)}$. This constraint follows from~\eqref{BoundaryConditions}.

\end{enumerate}

Physically, one might consider a black hole metric \eqref{newmetricGP2005} with an increasing rotation parameter~$a$, until extremality of the horizon is achieved. In such a scenario, the horizon extremality is associated with the rotation. However, non-rotating charged black holes might be extremal as well. In this case, the electric (magnetic) charge takes the role of the extremal parameter.

The situation is different with EIHs, since its five parameters can be chosen arbitrarily. The extremality is already assumed in our result \eqref{eq:MetricFunctionIH}, and therefore it can not be associated with just the electric charge.

In order to systematically analyze specific mutual relations between the two forms of the extremal black hole horizons, we will investigate three distinct subcases, namely:

\begin{enumerate}

\item Non-twisting black holes: ${a = 0}$, ${l = 0}$,

\item Non-rotating black holes: ${a = 0}$, ${l \neq 0}$,

\item General rotating black holes: ${a \neq 0}$ (with subcases ${l = 0}$, ${\alpha = 0}$, and ${e = 0 = g}$).

\end{enumerate}

\vspace{2mm}
\noindent
Each of these important cases will now be investigated in the following sections.

\subsection{Non-twisting black holes: ${a = 0}$, ${l = 0}$}
\label{sec:NonTwistingBlackHoles}

By setting \emph{both} the rotation parameter~$a$ and the NUT parameter~$l$ to zero, the null congruences corresponding to the (double degenerate) principal null directions \eqref{PDNPtetrad} become \emph{non-twisting}, see \cite{Griffiths2005,Griffiths2009}. In such a case, the metric \eqref{newmetricGP2005} considerably simplifies (setting ${l = 0}$ at first, then fixing the twist parameter ${\omega = a}$, and finally setting ${a = 0}$) to
\begin{align}
\dd s^2 = -\frac{1}{\Omega^2} &
  \left(-\frac{\QQ}{r^2}\,\dd t^2 + \frac{r^2}{\QQ}\,\dd r^2  +
  r^2 \Big( \,\frac{\dd\theta^2}{P} +  P\,\sin^2\theta\,\dd\fii^2\Big) \right),
  \label{newmetricGP2005a=0l=0}
\end{align}
where
\begin{align}
\Omega (r,\theta) & = 1 - \alpha\, r \cos\theta, \nonumber\\
P(\theta) & = (1-\alpha\, m\cos\theta)^2, \nonumber\\
\QQ (r)   & = (r-m)^2 (1-\alpha\,r) (1+\alpha\,r).
  \label{factorizedQa=0l=0}
\end{align}

In addition to two acceleration horizons at ${r_{{\rm a}\pm}=\pm \frac{1}{\alpha}}$, there is obviously the \emph{extremal black hole horizon} located at
\begin{equation}
r_H=m,  \qquad\hbox{where}\qquad  m^2=e^2+g^2, \label{eq:extremality-nontwist}
\end{equation}
see \eqref{extremeroot-alpha=0nebol=0}. Its area \eqref{eq:HorizonArea} reads
\begin{align}\label{eq:HorizonArea-nontwist}
A = \frac{ 4\pi C\, m^2} {1-\alpha^2m^2}.
\end{align}

The explicit metric \eqref{newmetricGP2005a=0l=0}, \eqref{factorizedQa=0l=0} describes \emph{extremely charged black holes which uniformly accelerate}. It is a generalization of extremal Reissner-Nordstr\"om spacetime to admit acceleration $\alpha$, and also the special (extremal) case ${e^2=m^2}$ of the charged C-metric (i.e. the acceleration extension of the metric (9.14), and the special extremal subcase of the metric (14.41) in \cite{Griffiths2009}).

Since the constants defined by \eqref{a34} are now ${ a_3 = 2\alpha \,m}$ and ${ a_4 = -\alpha^2 m^2}$, the \emph{deficit angles} \eqref{eq:DeficitAngles}
at the two poles on the extremal horizon are simply
\begin{align}
\begin{split}
\delta_+ &= 2\pi C (1 - \alpha \,m)^2 -2\pi, \\
\delta_- &= 2\pi C (1 + \alpha \,m)^2 -2\pi. \label{eq:DeficitAngles-nontwist}
\end{split}
\end{align}
By choosing the conicity parameter ${C=(1 - \alpha \,m)^{-2}}$ we can always achieve ${\delta_+ =0 }$, while by
${C=(1 + \alpha \,m)^{-2}}$ we obtain ${\delta_- =0}$. However, it is physically more important to evaluate the \emph{difference of these two deficit angles} at opposite poles of the extremal horizon,
\begin{align}
\Delta \equiv \delta_- -\delta_+ = 8\pi C \,\alpha \,m . \label{eq:DeficitAngleDifference-nontwist}
\end{align}
It exactly agrees with the interpretation of $\delta_{\pm}$ as parameters characterizing cosmic strings (or struts) with a tension (or compression) force resulting in the acceleration ${\alpha}$ to one side. If ${\alpha=0}$, there is an equilibrium between the tensions excerted by the two opposite strings and the black hole is static. Introduction of a Newtonian force
\begin{align}
F_N \equiv \frac{\delta_- - \delta_+}{8\pi C} = \alpha \,m , \label{eq:NewtonianForce}
\end{align}
makes this relation manifest.
For ${\alpha=0}$, ${C=1}$ we recover from  \eqref{newmetricGP2005a=0l=0} the usual extremal Reissner-Nordstr\"om spacetime without the conical singularities.

Now we employ the general expression \eqref{eq:MetricFunction} valid for all type~D black holes at their extremal horizon. For the particular metric \eqref{newmetricGP2005a=0l=0}, \eqref{factorizedQa=0l=0} we obtain, using the relations ${\rho^2 = r^2}$, ${\Omega = 1 - \alpha\, r\, \varsig}$ and
${\tilde{P} = (1 - \varsig^2)(1 - a_3\, \varsig - a_4\,\varsig^2) = (1 - \varsig^2)(1 - \alpha \,m\, \varsig)^2}$,  evaluated at ${r=r_H=m}$,  that
\begin{align*}
f_{\text{D}}(\varsig) &=  \frac{4\pi C^2\,m^2}{A}\,(1 - \varsig^2).
\end{align*}
Substituting form \eqref{eq:HorizonArea-nontwist} and \eqref{eq:ZetaAndVarsigma}, which simplifies
${\varsig = (\zeta + \alpha\, m)/(1 + \alpha\,m\,\zeta)}$, we get
\begin{align}
f_{\text{D}}(\zeta) &= C (1-\alpha^2m^2)^2 \,\frac{1 - \zeta^2}{(1 + \alpha\,m\, \zeta)^2}.
\label{eq:MetricFunction-nontwist}
\end{align}

This have to be compared with the metric function \eqref{eq:MetricFunctionIH} for the general axisymmetric extremal isolated horizon, expressed in the canonical coordinates. Using \eqref{eq:DeficitAngles-nontwist} we obtain the relations
\begin{align*}
& (2\pi+\delta_+)(2\pi+\delta_-) = 4\pi^2 C^2\, (1 - \alpha^2 m^2)^2, \\
& 4\pi(1 + \zeta^2)+ \delta_-(1 + \zeta)^2 +\delta_+(1-\zeta)^2 + q^2 (1-\zeta^2) \\
& \qquad = 8\pi C\,\Big[ \big[\tfrac{1}{2}(1 + \alpha^2m^2) + \tfrac{1}{8\pi C}\,q^2\big]
  + 2\alpha\,m\, \zeta
  + \big[\tfrac{1}{2}(1 + \alpha^2m^2) - \tfrac{1}{8\pi C}\,q^2\big]\,\zeta^2 \Big],
\end{align*}
so that
\begin{align} \label{eq:MetricFunctionIH-nontwist}
  f_{\text{EIH}}(\zeta) &=
  \frac{C\, (1 - \alpha^2 m^2)^2\,(1-\zeta^2)}
  {\big[\tfrac{1}{2}(1 + \alpha^2m^2) + \tfrac{1}{8\pi C}\,q^2\big]
  + 2\alpha\,m\, \zeta
  + \big[\tfrac{1}{2}(1 + \alpha^2m^2) - \tfrac{1}{8\pi C}\,q^2\big]\,\zeta^2}.
\end{align}
It is obvious that ${f_{\text{EIH}} = f_{\text{D}}}$ for \emph{all values} of $\zeta$ if, and only if,
${\tfrac{1}{2}(1 + \alpha^2m^2)=1-\tfrac{1}{8\pi C}\,q^2}$, or for the \emph{unique choice} of the dimensionless charge parameter
\begin{align}
q^2 & = 4\pi C\,( 1 - \alpha^2m^2 ).
\label{eq:UniqueRelation-nontwist}
\end{align}
We achieved a \emph{perfect agreement} between the metric function $f_{\text{EIH}}(\zeta)$ for the extremal isolated horizon and $f_{\text{D}}(\zeta)$ given by \eqref{eq:MetricFunction-nontwist} for non-twisting type~D black holes.

Moreover, by substituting \eqref{eq:Definiceq^2} with \eqref{eq:HorizonArea-nontwist} into \eqref{eq:UniqueRelation-nontwist} we obtain
\begin{align}
Q_E^2 + Q_M^2 & = C^2\, m^2 = C^2\, (e^2+g^2).
\label{eq:RelationBetweenCharges-nontwist}
\end{align}
This is an explicit relation between the \emph{total electric and magnetic charges}~$Q_E$ and~$Q_M$, determined by the Gauss law \eqref{eq:IHcharge}, and the charge parameters~$e$ and~$g$ in the black hole metric \eqref{newmetricGP2005a=0l=0}--\eqref{eq:extremality-nontwist}, which is the special case of type~D family \eqref{newmetricGP2005}. For ${C=1}$, these parameters match, ${Q_E=e}$ and ${Q_M=g}$.

Finally, notice also that the condition \eqref{eq:UniqueRelation-nontwist} together with \eqref{eq:DeficitAngles-nontwist} gives the constant $c_{\pi}$ real. Consequently, the function $\pinp$ is real, as well as $\Psi_2$. In general, an angular momentum of an IH is determined from the imaginary part of $\Psi_2$ \cite{Ashtekar2004-multipolemoments}. It is zero in this studied case, so the \emph{angular momentum vanishes}, in agreement with our assumption of
non-twisting black holes ${a = 0}$, ${l = 0}$.

\subsection{Non-rotating black holes: ${a = 0}$}
\label{sec:NonRotatingBlackHoles}

For ${a = 0}$, we should obtain a non-rotating and accelerating charged black hole with the NUT parameter~$l$. However, it was explicitly shown in \cite{Griffiths2005} that, in this setting, the acceleration parameter $\alpha$ is a pure gauge and can be eliminated by a coordinate transformation.

Interestingly, accelerating NUT black holes exist \cite{ChngMannStelea2006}, but they are \emph{not} included in the Pleba\'{n}ski-Demia\'{n}ski family of type~D metrics \eqref{PDmetric}. In a recent comprehensive study \cite{PodolskyVratny2020} it was demonstrated that these spacetimes are algebraically general (of type~I), and thus do not belong to the investigated family of type~D black holes. In any case, such accelerating NUT spacetimes do not admit extreme black hole horizons.

Hence, without loss of generality, we can set ${\alpha = 0}$, and the only non-zero parameters remain the mass $m$, the NUT parameter $l$, the charges $e, g$ and the free conicity constant~$C$. One of these parameters is  fixed by the extremality condition.

For ${a=0=\alpha}$ we obtain ${a_3=0=a_4}$ and ${\omega^2 k = -l^2}$, so that
the general type~D black hole metric \eqref{newmetricGP2005} reduces to charged NUT black hole \begin{align}
\dd s^2 =  &
  \,\frac{\QQ}{\rho^2}\Big(\dd t- 4l\sin^2\!\frac{\theta}{2}\,\dd\fii \Big)^2 - \frac{\rho^2}{\QQ}\,\dd r^2 - \rho^2\,(\dd\theta^2 + \sin^2\theta\,\dd\fii^2),
  \label{chargedNUTmetric}
\end{align}
where ${\rho^2 = r^2 + l^2}$ and ${\QQ (r) = (e^2+g^2-l^2) -2m\,r + r^2}$.

It follows from \eqref{eq:HorizonArea} that the area of the corresponding horizon is
\begin{align}
A = 4\pi C\, (r_H^2+l^2), \label{eq:AzeroorrHF}
\end{align}
so that the value $r_H$ of radial coordinate on the horizon is ${r_H^2 = \frac{1}{4\pi C}\,A - l^2}$.
Substitution into \eqref{eq:MetricFunction} (with ${\tilde{P} = 1 - \varsig^2}$, ${\Omega=1}$, ${\rho^2 = r_H^2+l^2}$ and ${\varsig=\zeta}$) reveals that
\begin{align}
f_{\text{D}}(\zeta) = C\,(1 - \zeta^2), \label{eq:RegularMetricFunction}
\end{align}
which (irrespective of ${l \ne 0}$) agrees with the expression \eqref{eq:MetricFunction-nontwist} for ${\alpha=0}$. This is the metric function of a quasi-regular sphere (regular provided ${C=1}$),
see Eq.~\eqref{CanonicalMetricSphere}. The NUT parameter, as well as the charges, thus preserve the spherical symmetry of the horizon.

Now we employ the \emph{extremality condition}~\eqref{eq:ConditionOfExtremality}. In the special case of vanishing acceleration, it leads to the relations \eqref{extremeroot-alpha=0nebol=0}, and for $a=0$ we obtain
\begin{align}\label{eq:ConditionOfExtremality-nonrot}
r_H=m,  \qquad\hbox{where}\qquad  m^2+l^2=e^2+g^2.
\end{align}
Using \eqref{eq:AzeroorrHF}, we can thus write
\begin{align}
A = 4 \pi C \,(m^2+l^2) = 4 \pi C \,(e^2 + g^2). \label{eq:ConditionOfExtremalityForAzero}
\end{align}
Moreover, the metric function $\QQ$ factorizes as
\begin{align}
\QQ (r) = (r-m)^2, \label{QchargedNUTextreme}
\end{align}
in the same way as for the extremal Reissner-Nordstr\"om spacetime, simplifying \eqref{chargedNUTmetric} to the metric of \emph{extremely charged NUT black holes}
\begin{align}\label{eq:ExtremalyChargedNUTmetric}
\dd s^2 =  &
  \frac{(r-m)^2}{r^2 + l^2}\Big(\dd t- 4l\sin^2\!\frac{\theta}{2}\,\dd\fii \Big)^2
  - \frac{r^2 + l^2}{(r-m)^2}\,\dd r^2   - (r^2 + l^2)\,(\dd\theta^2 + \sin^2\theta\,\dd\fii^2).
\end{align}

The \emph{deficit angles} on the extremal horizon are given by expressions
\eqref{eq:DeficitAngles}.\footnote{Or by \eqref{eq:DeficitAnglesBar}, yielding ${\bar\delta_+ = \bar\delta_-= 2\pi\, (C-1)}$ in the alternative metric form (\ref{PDmetricGPbar}).} As before, ${a_3=0=a_4}$ and they reduce to
\begin{align}
\delta_{+} = \delta_{-} = 2\pi\, (C-1).
 \label{deficitanglesNUTextreme}
\end{align}
For ${C=1}$, both the poles are regular (and since ${\Delta\equiv \delta_- -\delta_+ =0}$, the black hole cannot accelerate).

In order to compare \eqref{eq:RegularMetricFunction} with the canonical metric function $f_{\text{EIH}} (\zeta)$ for extremal isolated horizons, we substitute  \eqref{deficitanglesNUTextreme} into \eqref{eq:MetricFunctionIH}, which yields
\begin{align}\label{eq:MetricFunctionIH-nonrot}
f_{\text{EIH}} (\zeta) = \frac{8 \pi C^2\, (1 - \zeta^2)}{4 \pi C\, (1 + \zeta^2)+q^2\, (1 - \zeta^2)}.
\end{align}
This function implicitly assumes extremality, which relates the area and the charges of the type~D black holes via \eqref{eq:ConditionOfExtremalityForAzero}. It remains to find the unique value of $q^2$ such that $f_{\text{EIH}}$ becomes the metric function $f_{\text{D}}$ of the form \eqref{eq:RegularMetricFunction}. Indeed, such $q^2$ exists, namely
\begin{align}\label{eq:RelationBetweenCharges-nonrot}
q^2 = 4\pi C \quad \Leftrightarrow \quad Q_E^2 + Q_M^2 = C^2\, (e^2 + g^2),
\end{align}
in full agreement with the previous relations \eqref{eq:UniqueRelation-nontwist},  \eqref{eq:RelationBetweenCharges-nontwist}.

We have thus completely identified the metric function $f_{\text{EIH}}$ of extremal isolated horizons with the corresponding expression $f_{\text{D}}$ for extremely charged NUT black holes \eqref{eq:ExtremalyChargedNUTmetric} of algebraic type~D, which are the only such non-rotating solutions (that is for ${a=0}$). Moreover, we have clarified explicit relations between the physical parameters ${(m, l, e, g, C)}$, constrained by~\eqref{eq:ConditionOfExtremality-nonrot}, and the dimensionless (geometric) parameters ${(\delta_+, \delta_-, q^2)}$ which enter the EIH metric function~\eqref{eq:MetricFunctionIH}. Interestingly, in view of  \eqref{deficitanglesNUTextreme}, \eqref{eq:RelationBetweenCharges-nonrot}, they are fully determined just by \emph{the single conicity parameter}~$C$.

It is also interesting to notice that the area $A$ of the extremal horizon, given by \eqref{eq:ConditionOfExtremalityForAzero}, is proportional to the ``effective squared mass'' ${m^2+l^2}$, which is equal to the ``effective squared charge'' ${e^2+g^2}$. Considering just the induced metric of the extremal horizon, described by the function $f_{\text{EIH}} \equiv f_{\text{D}}$, we can not distinguish between the separate contribution from the mass parameter~$m$ and the NUT parameter~$l$ (or, equivalently, the electric charge~$e$ and the magnetic charge~$g$).

\subsection{Rotating black holes: ${a \neq 0}$}
\label{sec:RotatingBlackHoles}

Finally, let us investigate the general case of type~D black holes of mass $m$ with rotation represented by the parameter $a$. This is a wide family of spacetimes, which contains the Kerr solution and its generalizations to include electromagnetic charges $e, g$, the NUT parameter $l$, and also the acceleration $\alpha$. In such a case, it is most natural to relate the extremality of the black hole to the rotation parameter~$a$, keeping the remaining physical parameters free.

For these generic black holes, the horizon extremality condition is \eqref{extremerootofQ}. It is convenient to employ the gauge \eqref{eq:ChoiceOfOmega} of the twist parameter~$\omega$ which implies \eqref{eq:ChoiceOfOmega-al}, i.e.,
\begin{align}\label{eq:a/omega-l/omega}
\frac{a}{\omega} = \frac{a^2}{a^2 + l^2}, \qquad
\frac{l}{\omega} = \frac{a\,l}{a^2 + l^2} .
\end{align}
Consequently, expressions \eqref{k} and \eqref{keg} can be written as
\begin{align}
  \frac{\omega^2 k}{a^2-l^2} &= \frac{1 + 2\A\,m - 3\A^2(e^2+g^2)}{1 + 3\A^2(a^2-l^2)}, \label{eq:omegak}\\
  \omega^2 k+e^2+g^2 &= \frac{(a^2-l^2+e^2+g^2) + 2\A\,m\,(a^2-l^2)} {1 + 3 \A^2 (a^2-l^2)}, \label{eq:omegak+e+g}
\end{align}
where we have introduced a unique combination of the three physical parameters
\begin{align}\label{eq:DefA}
\A \equiv  \frac{\alpha\,a\,l}{a^2 + l^2}.
\end{align}
Then, the \emph{generic extremality condition} reads
\begin{equation}
 \big[\, m - \A\, (\omega^2k+e^2+g^2)\big]^2=\frac{\omega^2k}{a^2-l^2}(\omega^2k+e^2+g^2),
  \label{extremerootofQgauge}
\end{equation}
and the corresponding (degenerate) \emph{extreme horizon is located at}
\begin{equation}
r_H=\big[\, m - \A\, (\omega^2k+e^2+g^2)\big]{\frac{a^2-l^2}{\omega^2k}},
  \label{extremeHorizonGeneral}
\end{equation}
or equivalently
\begin{equation}
r_H=\frac{(\omega^2k+e^2+g^2)}{m - \A\, (\omega^2k+e^2+g^2)}.
  \label{extremeHorizonGeneral2}
\end{equation}

By substituting from \eqref{eq:omegak}, \eqref{eq:omegak+e+g} we obtain explicit expressions for the horizon
\begin{equation}
r_H=\frac{m-\A\,(a^2-l^2+e^2+g^2)+\A^2\, m\,(a^2-l^2)}{1+2\A\, m - 3\A^2(e^2+g^2)},
  \label{extremeHorizonGeneralExpl}
\end{equation}
or equivalently
\begin{equation}
r_H=\frac{(a^2-l^2+e^2+g^2) + 2\A\,m\,(a^2-l^2)}{m - \A\, (a^2-l^2+e^2+g^2) + \A^2\,m\,(a^2-l^2)},
  \label{extremeHorizonGeneralExpl2}
\end{equation}
so that
\begin{equation}
r_H^2=\frac{(a^2-l^2+e^2+g^2) + 2\A\,m\,(a^2-l^2)}{1+2\A\, m - 3\A^2(e^2+g^2)}.
  \label{extremeHorizonGeneralExplSquare}
\end{equation}

For ${\A=0}$, which occurs whenever ${\alpha=0}$ or ${a=0}$ or ${l=0}$, these expressions simplify considerably to
\begin{equation}
r_H =  m,  \qquad\hbox{where}\qquad   m^2=a^2-l^2+e^2+g^2,
  \label{extremeHorizonGeneralA=0}
\end{equation}
in full agreement with the previously investigated cases ${a=0}$, ${l=0}$, see \eqref{eq:extremality-nontwist}  and \eqref{eq:ConditionOfExtremality-nonrot}.

\subsubsection{Accelerating Kerr-Newman black holes: $l=0$}
\label{sec:AccKerrNew}

In this section, we assume a vanishing NUT parameter $l$, so that $\A$ given by \eqref{eq:DefA} is zero. In such a case,
\begin{align}
r_H = m,  \qquad\hbox{where}\qquad   m^2=a^2+e^2+g^2.
  \label{extremerootofQgaugel=0}
\end{align}
see \eqref{extremeHorizonGeneralA=0} and also \eqref{extremeroot-alpha=0nebol=0}. Possible values of the extremal rotation parameter are thus
\begin{align}
a = \pm \sqrt{m^2 - (e^2+g^2)},\label{eq:ExtremalRotationParameterForlZero}
\end{align}
where the two signs describe two opposite orientations of the black hole rotation. Without loss of generality we may choose the positive one.

Such a family of spacetimes represents \emph{accelerating extremal Kerr-Newman black holes}. Using \eqref{eq:a/omega-l/omega}-\eqref{eq:omegak+e+g} with ${\A=0}$, implying ${a_3 = 2\alpha \,m}$ and ${a_4 = -\alpha^2 m^2}$, its metric \eqref{newmetricGP2005} reads
\begin{align}
\dd s^2 = -\frac{1}{\Omega^2} &
  \left(-\frac{\QQ}{\rho^2}\left[\dd t- a\sin^2\theta \,\dd\fii \right]^2 + \frac{\rho^2}{\QQ}\,\dd r^2 \right. \nonumber\\
& \quad \left. + \,\frac{\rho^2}{P}\,\dd\theta^2
  + \frac{P}{\rho^2}\,\sin^2\theta\, \big[ a\dd t - (r^2+a^2)\dd\fii \big]^2
 \right), \label{newmetricGP2005l=0}
\end{align}
where
\begin{align}
\Omega (r,\theta) &= 1 - \alpha\, r \cos\theta , \nonumber\\
\rho^2 (r,\theta) &= r^2 + a^2 \cos^2\!\theta, \nonumber\\
P(\theta) &= (1-\alpha\, m\cos\theta)^2 , \nonumber\\
\QQ (r) &= (r-m)^2 (1-\alpha\,r)(1+\alpha\,r).
  \label{factorizedQl=0}
\end{align}
For ${a=0}$ we recover Eqs.~\eqref{newmetricGP2005a=0l=0}, \eqref{factorizedQa=0l=0}.

The corresponding metric function $f_{\text{D}}(\zeta)$ given by \eqref{eq:MetricFunction} describes the geometry of the extremal horizon located at ${r_H = m}$ with the area
\begin{align}\label{eq:HorizonAreal=0}
A = 4\pi C\, \frac{r_H^2+a^2}{1-\alpha^2 m^2 }
  = 4 \pi C \, \frac{2m^2 - (e^2 + g^2)}{\left(1 + \alpha m\right)\left(1 - \alpha m\right)}.
\end{align}
In terms of the coordinate $\varsig$, the metric function of the horizon reads
\begin{align*}
f_{\text{D}}(\varsigma) &= C\, (1-\alpha^2 m^2)(m^2+a^2)\,
    \frac{1 - \varsig^2}{m^2 + a^2\, \varsig^2}.
\end{align*}
The transformation \eqref{eq:ZetaAndVarsigma} is
${\varsig = (\zeta + \alpha\,m)/(1 + \alpha\,m\,\zeta)}$,
so that
\begin{align}
f_{\text{D}}(\zeta) &= C\, (1-\alpha^2 m^2)^2\,
    \frac{ (m^2+a^2)(1-\zeta^2)}{m^2(1 + \alpha\,m\,\zeta)^2 + a^2\,(\zeta + \alpha\,m)^2},
\label{eq:MetricFunctionD-NoNUT}
\end{align}
which is equivalent to
\begin{align*} 
f_{\text{D}}(\zeta) &= \frac{ C\,(1-\alpha^2 m^2)^2\,(2m^2 - e^2 - g^2)(1-\zeta^2)}
{m^2 (1 + \alpha^2\,m^2)(1 + \zeta^2) + 4 \alpha\,m^3\,\zeta - (e^2 + g^2)(\zeta + \alpha m)^2}.
\end{align*}

Now, the deficit angles \eqref{eq:DeficitAngles} at the two poles of the horizon are
\begin{align}
\begin{split}
\delta_+ &= 2\pi C (1 - \alpha \,m)^2 -2\pi, \\
\delta_- &= 2\pi C (1 + \alpha \,m)^2 -2\pi, \label{eq:DeficitAnglesNoNUT}
\end{split}
\end{align}
which are the same relations as in the non-twisting case \eqref{eq:DeficitAngles-nontwist}. Therefore, the metric function \eqref{eq:MetricFunctionIH} for the extremal isolated horizon in the canonical coordinates must have the same form as \eqref{eq:MetricFunctionIH-nontwist}, i.e.
\begin{align} \label{eq:MetricFunctionIH-NoNUT}
  f_{\text{EIH}}(\zeta) &=
  \frac{C\, (1 - \alpha^2 m^2)^2\,(1-\zeta^2)}
  {\big[\tfrac{1}{2}(1 + \alpha^2m^2) + \tfrac{1}{8\pi C}\,q^2\big]
  + 2\alpha\,m\, \zeta
  + \big[\tfrac{1}{2}(1 + \alpha^2m^2) - \tfrac{1}{8\pi C}\,q^2\big]\,\zeta^2}.
\end{align}
We achieve an agreement between the results \eqref{eq:MetricFunctionD-NoNUT} and \eqref{eq:MetricFunctionIH-NoNUT} by a unique choice of the dimensionless charge parameter~$q$ as
\begin{align}
q^2 & = 4\pi C\,( 1 - \alpha^2m^2 )\,\frac{r_H^2-a^2}{r_H^2+a^2}  = 4\pi C\,( 1 - \alpha^2m^2 )\,\frac{e^2 + g^2}{2m^2 - e^2 - g^2}.
\label{eq:UniqueRelation-NoNUT}
\end{align}
Interestingly, it is formally possible to set ${a=0}$, recovering the previous relation \eqref{eq:UniqueRelation-nontwist}.

Again, we have achieved a \emph{perfect agreement} between the metric function $f_{\text{EIH}}(\zeta)$ for the extremal isolated horizon and $f_{\text{D}}(\zeta)$ given by \eqref{eq:MetricFunctionD-NoNUT} for type~D black holes without the NUT parameter~$l$, that is for the whole family of accelerating Kerr-Newman black holes.

Finally, by substituting \eqref{eq:Definiceq^2} with \eqref{eq:HorizonAreal=0} into \eqref{eq:UniqueRelation-NoNUT} we obtain the same simple expression as in the non-rotating and non-twisting cases
\begin{align}
Q_E^2 + Q_M^2 & = C^2\, (m^2-a^2) = C^2\, (e^2+g^2),
\label{eq:RelationBetweenCharges-NoNUT}
\end{align}
relating the total (physically defined) electric and magnetic charges~$Q_E$ and~$Q_M$ and the charge parameters~$e$ and~$g$ in the metric \eqref{newmetricGP2005a=0l=0}--\eqref{eq:extremality-nontwist}. Clearly, ${Q_E=e}$ and ${Q_M=g}$ for ${C=1}$, in which case the acceleration is caused by the two cosmic strings with the deficit angles
${\delta_+ = 2\pi\,\alpha \,m (\alpha\,m-2) }$,
${\delta_- = 2\pi\,\alpha \,m (\alpha \,m+2) }$.
For ${\alpha=0}$, there is no acceleration and we recover the stationary extreme Kerr-Newman black hole.

\subsubsection{Kerr-Newman-NUT black holes: ${\alpha = 0}$}
\label{sec:KerrNewmanNUTBlackHoles}

For vanishing acceleration ${\alpha=0}$, which implies ${\A=0}$, ${a_3 = 0= a_4}$, ${\Omega = 1}$, ${P=1}$, the position of the extremal horizon and the corresponding extremality condition have the form \eqref{extremeHorizonGeneralA=0}, namely
\begin{equation}
r_H = m, \qquad\hbox{where}\qquad m^2=a^2-l^2+e^2+g^2.
  \label{extremerootofQgaugealpha=0}
\end{equation}
Its area \eqref{eq:HorizonArea} is
\begin{align}\label{eq:HorizonArea-alpha=0}
A = 4\pi C\, [r_H^2+(a+l)^2].
\end{align}
In this case, the generic black hole metric \eqref{newmetricGP2005}  simplifies to
\begin{align}
\dd s^2 =  &
  \frac{\QQ}{\rho^2}\left[\dd t- \left(a\sin^2\theta +4l\sin^2\!\frac{\theta}{2} \right)\dd\fii \right]^2 - \frac{\rho^2}{\QQ}\,\dd r^2  \nonumber\\
& \quad  - \,\rho^2\dd\theta^2
  - \frac{\sin^2\theta}{\rho^2}\, \Big[ a\dd t -\big(r^2+(a+l)^2\big)\,\dd\fii \Big]^2, \label{newmetricGP2005alpha=0}
\end{align}
where
\begin{align}
\rho^2 (r,\theta) &= r^2 + (l + a \cos\theta)^2, \nonumber\\
\QQ (r) &= (r-m)^2 .
  \label{factorizedQalpha=0}
\end{align}
This is a family of stationary (non-accelerating) \emph{extremal Kerr-Newman-NUT black holes}.
In a correspondence with the previous results, the metric of extremely charged NUT black holes \eqref{eq:ExtremalyChargedNUTmetric} is obtained by setting $a = 0$.

The deficit angles \eqref{eq:DeficitAngles} at the two poles of the horizon are
\begin{align}
\delta_+ = 2\pi \big(C - 1 \big),\qquad
\delta_- = 2\pi \Big(C\,\frac{r_H^2+(a+l)^2}{r_H^2+(a-l)^2} - 1 \Big),
\label{eq:DeficitAnglesalpha=0}
\end{align}
where the values of the parameters are constrained by \eqref{extremerootofQgaugealpha=0}. The first pole is regular for ${C=1}$. The deficit angles of the alternative metric form \eqref{PDmetricGPbar} are
\begin{align}
\bar\delta_+ = 2\pi \Big(C\,\frac{r_H^2+(a-l)^2}{r_H^2+(a+l)^2} - 1 \Big), \qquad
\bar\delta_- = 2\pi \big(C - 1 \big),
\label{eq:DeficitAnglesBaralpha=0}
\end{align}
and thus the second pole is regular for ${C=1}$.

From \eqref{eq:MetricFunction}, using the relation ${\varsig=\zeta}$ which follows from \eqref{eq:ZetaAndVarsigma}, we now obtain
\begin{align}
f_{\text{D}}(\zeta) &= C\,\big[r_H^2+(a + l)^2\big]\,
    \frac{1 - \zeta^2}{r_H^2 + (l + a\, \zeta)^2}.
\label{eq:MetricFunction-alpha=0}
\end{align}

This should correspond to the canonical metric function $f_{\text{EIH}} (\zeta)$ of the extremal isolated horizons \eqref{eq:MetricFunctionIH}, whose denominator can be rewritten as
\begin{align}
  & 4\pi(1 + \zeta^2)+ \delta_-(1 + \zeta)^2 +\delta_+(1-\zeta)^2 + q^2 (1-\zeta^2)  \nonumber\\
  &\qquad = [(\delta_+ + \delta_- + 4\pi)+q^2] + 2 (\delta_- - \delta_+)\,\zeta+ [(\delta_+ + \delta_- + 4\pi)-q^2]\,\zeta^2. \label{eq:MetricFunctionIHdenominator}
\end{align}
Substituting from \eqref{eq:DeficitAnglesalpha=0},
which implies
\begin{align*}
(\delta_+ + \delta_- + 4\pi)\,\big[{r_H^2+(a-l)^2}\big] &= 4\pi C\,\big[r_H^2+(a^2+l^2)\big],\\
(\delta_- - \delta_+)\,\big[{r_H^2+(a-l)^2}\big] &=  8\pi C\, al,
\end{align*}
we thus obtain
\begin{align} \label{eq:MetricFunctionIH-alpha=0}
  f_{\text{EIH}}(\zeta) &= C\, \big[r_H^2+(a+l)^2\big]\,
     \frac{1-\zeta^2}{c_0 + 2 al\,\zeta + c_2\,\zeta^2},
\end{align}
where
\begin{align*}
c_0 &= \frac{1}{2}\big[r_H^2+(a^2+l^2)\big] + \frac{q^2}{8\pi C}\big[r_H^2+(a-l)^2\big],\\
c_2 &= \frac{1}{2}\big[r_H^2+(a^2+l^2)\big] - \frac{q^2}{8\pi C}\big[r_H^2+(a-l)^2\big].
\end{align*}
Clearly, for the \emph{unique value} of the dimensionless charge parameter
\begin{align}
q^2 & = 4\pi C\,\frac{r_H^2-a^2+l^2}{r_H^2+(a-l)^2},
\label{eq:UniqueRelation-alpha=0}
\end{align}
we get ${c_0 = r_H^2+l^2}$ and ${c_2 = a^2}$, so that ${c_0 + 2 al\,\zeta + c_2\,\zeta^2 = r_H^2+(l + a\,\zeta)^2}$. The  metric function \eqref{eq:MetricFunctionIH-alpha=0} thus \emph{fully agrees} with $f_{\text{D}}(\zeta)$ given by \eqref{eq:MetricFunction-alpha=0} of all non-accelerating type~D black holes.

Notice that for ${a=0}$ the function ${f_{\text{EIH}}\equiv f_{\text{D}}}$ formally reduces to ${C\,(1 - \zeta^2)}$, which is Eq.~\eqref{eq:RegularMetricFunction}, while \eqref{eq:UniqueRelation-alpha=0} simplifies to ${q^2 = 4\pi C}$,  which is Eq.~\eqref{eq:RelationBetweenCharges-nonrot}.
Similarly, for ${l=0}$ we obtain
\begin{align*}
f_{\text{EIH}}\equiv f_{\text{D}} = C\,\frac{ (r_H^2+a^2)(1-\zeta^2)}{r_H^2 + a^2\,\zeta^2}, \qquad
q^2  = 4\pi C\,\frac{r_H^2-a^2}{r_H^2+a^2},
\end{align*}
which are the expressions \eqref{eq:MetricFunctionD-NoNUT} and \eqref{eq:UniqueRelation-NoNUT}, respectively, when ${\alpha=0}$.

Moreover, from the relations \eqref{eq:Definiceq^2} and \eqref{eq:UniqueRelation-alpha=0}, together with \eqref{eq:HorizonArea-alpha=0} and the extremality condition \eqref{extremerootofQgaugealpha=0}, it follows that
\begin{align} \label{eq:RelationBetweenCharges-alpha=0}
Q_E^2 + Q_M^2 &= C^2\, \frac{r_H^2+(a+l)^2}{r_H^2+(a-l)^2}\, (r_H^2-a^2+l^2) \nonumber\\
 &\equiv C^2\, \frac{m^2+(a+l)^2}{m^2+(a-l)^2}\, (e^2+g^2).
\end{align}
With this specific rescaling, the parameters $e$ and $g$ give the \emph{genuine physical electric and magnetic charges} $Q_E$ and $Q_M$, respectively.

Interestingly, \emph{whenever either ${a=0}$ or ${l=0}$}, this relation simplifies to
\begin{align}
Q_E^2 + Q_M^2 = C^2\, (e^2 + g^2),
\end{align}
recovering \eqref{eq:RelationBetweenCharges-nontwist}, \eqref{eq:RelationBetweenCharges-nonrot} and \eqref{eq:RelationBetweenCharges-NoNUT}.

Of course, extremal rotating vacuum black holes \emph{without the electromagnetic field} are obtained simply by setting ${e=0=g}$.

\subsubsection{General black holes of algebraic type~D with extremal horizon}
\label{sec:GeneralBlackHoles}

A generic case contains 6 physical parameters, namely $m$, $e$, $g$, $a$, $l$, $\alpha$, plus the conicity parameter $C$, which are constrained by the extremality condition \eqref{extremerootofQgauge} of the horizon at $r_H$ given by the formula \eqref{extremeHorizonGeneralExpl}. At first glance, it seems too complicated to prove analytically the equivalence of the functions $f_{\text{D}}(\zeta)$ and $f_{\text{EIH}}(\zeta)$ given by the expressions \eqref{eq:MetricFunction} and \eqref{eq:MetricFunctionIH}, respectively. We will now demonstrate that even this most general case is explicitly solvable.

The key point is to realize that the constants $a_3$ and $a_4$ defined by \eqref{a34} --- which directly determine the metric function $\tilde{P}(\zeta)$ via \eqref{eq:PolynomialPtilde} and also the deficit angles \eqref{eq:DeficitAngles} --- are \emph{mutually related due to the extremality condition} \eqref{extremerootofQgauge} as
\begin{align}
a_3^2=-4\,a_4. \label{eq:a3a4extremal}
\end{align}
Indeed,
\begin{align}
a_3 &= \frac{2\alpha a^2}{a^2+l^2}\,\big[\,m -2\A\,(\omega^2k+e^2+g^2)\big],  \nonumber\\
a_4 &= -\frac{\alpha^2 a^4}{(a^2+l^2)^2}\,(\omega^2k+e^2+g^2), \label{a34A}
\end{align}
so that
\begin{align*}
a_3^2+4\,a_4 &= \frac{4\alpha^2 a^4}{(a^2+l^2)^2}\,\Big[[m -2\A\,(\omega^2k+e^2+g^2)]^2-(\omega^2k+e^2+g^2)\Big]  \nonumber\\
     &\equiv \frac{4\alpha^2 a^4}{(a^2+l^2)^2}\,\Big[[m -\A\,(\omega^2k+e^2+g^2)]^2 \nonumber\\
     &\hspace{20mm} -(\omega^2k+e^2+g^2)[1+2\A\, m -3\A^2(\omega^2k+e^2+g^2)]\Big] .\nonumber
\end{align*}
Applying the extremality condition \eqref{extremerootofQgauge}, we obtain
\begin{align}
a_3^2+4\,a_4 = \frac{4\alpha^2 a^4}{(a^2+l^2)^2}\,(\omega^2k+e^2+g^2)\,\Big[\, \frac{\omega^2k}{a^2-l^2}  -(1+2\A\, m) + 3\A^2(\omega^2k+e^2+g^2)\Big] ,\label{a34AA}
\end{align}
and after substituting from \eqref{eq:omegak}, \eqref{eq:omegak+e+g} we get ${a_3^2+4\,a_4=0}$.

This unique relation has important consequences in simplifying the key expressions. First, the metric function
$\tilde{P}(\zeta)$ given by  \eqref{eq:PolynomialPtilde} \emph{factorizes as}
\begin{align}
\tilde{P}(\varsig) = (1 - \varsig^2)(1 - \tfrac{1}{2}  a_3\,\varsig)^2. \label{eq:extremalP}
\end{align}
Moreover,
\begin{align}
\begin{split}
(1 - a_3 - a_4) &= (1 - \tfrac{1}{2}  a_3)^2,\\
(1 + a_3 - a_4) &= (1 + \tfrac{1}{2}  a_3)^2,\\
(1 - a_3 - a_4)(1 + a_3 - a_4) &= (1 + a_4)^2,
\label{eq:a3a4identities}
\end{split}
\end{align}
so that the explicit expressions \eqref{eq:DeficitAngles} for the deficit angles also simplify. In particular,
\begin{align}
(2\pi+\delta_-)(2\pi+\delta_+)\,d_- &= 4\pi^2 C^2\, (1 + a_4)^2\,d_+, \nonumber\\
(\delta_+ + \delta_- + 4\pi)\,d_- &= 2\pi C\,\big[(1 + \tfrac{1}{2}  a_3)^2\,d_+ + (1 - \tfrac{1}{2}  a_3)^2\,d_-\big], \label{eq:extremalDeficitAngles}\\
(\delta_- - \delta_+)\,d_- &=  2\pi C\,\big[(1 + \tfrac{1}{2}  a_3)^2\,d_+ - (1 - \tfrac{1}{2}  a_3)^2\,d_-\big],\nonumber
\end{align}
where
\begin{align}
d_+ \equiv r_H^2+(a+l)^2, \qquad
d_- \equiv r_H^2+(a-l)^2. \label{eq:d+-}
\end{align}
These expressions \emph{directly determine the function} $f_{\text{EIH}}(\zeta)$ given by \eqref{eq:MetricFunctionIH}. Rewriting its denominator as \eqref{eq:MetricFunctionIHdenominator}, with \eqref{eq:extremalDeficitAngles} we obtain
\begin{align} \label{eq:MetricFunctionEIH}
  f_{\text{EIH}}(\zeta) &= C\,d_+\, (1 + a_4)^2\,  \frac{1-\zeta^2}{ c_0 + c_1 \,\zeta + c_2\,\zeta^2},
\end{align}
in which
\begin{align}
c_0 & \equiv \frac{1}{4}\, \Big[(1 + \tfrac{1}{2}  a_3)^2\,d_+ + (1 - \tfrac{1}{2}  a_3)^2\,d_-\Big]
      + \frac{d_-}{8\pi C}\,q^2 , \nonumber\\
c_1 & \equiv \frac{1}{2}\, \Big[(1 + \tfrac{1}{2}  a_3)^2\,d_+ - (1 - \tfrac{1}{2}  a_3)^2\,d_-\Big],  \label{eq:c0c1c2}\\
c_2 & \equiv \frac{1}{4}\, \Big[(1 + \tfrac{1}{2}  a_3)^2\,d_+ + (1 - \tfrac{1}{2}  a_3)^2\,d_-\Big]
      - \frac{d_-}{8\pi C}\,q^2. \nonumber
\end{align}

This should now be compared with the function $f_{\text{D}}(\zeta)$ representing the metric on the extremal horizon of the most general black hole of algebraic type~D, as given by the equation \eqref{eq:MetricFunction}. Due to the nice factorization \eqref{eq:extremalP}, this function takes the form
\begin{align}
f_{\text{D}}(\varsig) &= \frac{4\pi C^2}{A} \,d_+^2\,\frac{(1 - \tfrac{1}{2}  a_3\,\varsig)^2(1 - \varsig^2)}
{\Omega^2(\varsig)\,\rho^2(\varsig)},
\label{eq:MetricFunctionFact}
\end{align}
where, using \eqref{eq:a/omega-l/omega}, the horizon area \eqref{eq:HorizonArea} is
\begin{align}\label{eq:HorizonAreaOmega}
A = \frac{ 4\pi C\,d_+\,(a^2 + l^2)^2}{(a^2 + l^2-\alpha\,a\,l\,r_H)^2-\alpha^2a^4\,r_H^2},
\end{align}
and\begin{align}
\Omega(\varsig) = \frac{(a^2 + l^2 - \alpha\,a\,l\,r_H) - \alpha\,a^2\,r_H\, \varsig}{a^2 + l^2}, \qquad
\rho^2(\varsig) = r_H^2 + (l + a\, \varsig)^2.  \label{eq:Omega/rho}
\end{align}

These functions have to be expressed in terms of the variable $\zeta$ via \eqref{eq:ZetaAndVarsigma}, that is
\begin{align}
\varsig = \frac{ (a^2 + l^2 - \alpha\,a\,l\,r_H )\,\zeta + \alpha\,a^2\,r_H}
{(a^2 + l^2 - \alpha\,a\,l\,r_H) + \alpha\,a^2\,r_H\,\zeta }, \label{eq:ZetaAndVarsigmaOmega}
\end{align}
which implies
\begin{align}
\Omega(\zeta)  & = \frac{1}{(a^2 + l^2)}\,\frac{(a^2 + l^2-\alpha\,a\,l\,r_H)^2-\alpha^2a^4\,r_H^2}
{(a^2 + l^2 - \alpha\,a\,l\,r_H) + \alpha\,a^2\,r_H\,\zeta}, \nonumber\\
\rho^2(\zeta) & = \frac{ C_0 + C_1 \,\zeta + C_2\,\zeta^2 }
{\big[(a^2 + l^2 - \alpha\,a\,l\,r_H) + \alpha\,a^2\,r_H\,\zeta \big]^2}, \nonumber\\
(1 - \varsig^2) & = \frac{ (a^2 + l^2-\alpha\,a\,l\,r_H)^2-\alpha^2a^4\,r_H^2 }
{\big[(a^2 + l^2 - \alpha\,a\,l\,r_H) + \alpha\,a^2\,r_H\,\zeta \big]^2}\,(1 - \zeta^2),  \label{eq:1-Varsigma^2}\\
(1 - \tfrac{1}{2}  a_3\,\varsig) & = \frac{ \big[(a^2 + l^2-\alpha\,a\,l\,r_H)-\frac{1}{2}a_3\,\alpha\,a^2\,r_H \big]
+ \big[\alpha\,a^2\,r_H -\frac{1}{2}a_3\,(a^2 + l^2-\alpha\,a\,l\,r_H)\big]\,\zeta}
{(a^2 + l^2 - \alpha\,a\,l\,r_H) + \alpha\,a^2\,r_H\,\zeta },  \nonumber
\end{align}
where
\begin{align}
C_0 & \equiv (r_H^2 + a^2 + l^2)(a^2 + l^2-\alpha\,a\,l\,r_H)^2 +\alpha^2a^4\,r_H^2 (a^2+l^2) \nonumber\\
    & \qquad - a^2(a^2+l^2)^2 + 4\alpha\,a^3l\, r_H (a^2 + l^2-\alpha\,a\,l\,r_H) , \nonumber\\
C_1 & \equiv 2a\,l \big[ (a^2 + l^2-\alpha\,a\,l\,r_H)^2+\alpha^2a^4\,r_H^2 \big] \label{eq:C0C1C2}\\
    & \qquad + 2\alpha\,a^2 r_H (r_H^2 + a^2 + l^2)(a^2 + l^2-\alpha\,a\,l\,r_H),  \nonumber\\
C_2 & \equiv a^2(a^2+l^2)^2 + \alpha^2a^4\,r_H^4.  \nonumber
\end{align}

Moreover, \emph{the constants $a_3$ and $a_4$}, related via \eqref{eq:a3a4extremal}, have to take the explicit form
\begin{align}
a_3  = \frac{2\alpha\,a^2\,r_H}{a^2 + l^2 - \alpha\,a\,l\,r_H}, \qquad
a_4  = -\frac{\alpha^2\,a^4\,r_H^2}{(a^2 + l^2 - \alpha\,a\,l\,r_H)^2},  \label{eq:a3a4explicit}
\end{align}
so that the function ${(1 - \tfrac{1}{2}  a_3\,\varsig)}$ simplifies to
\begin{align}
(1 - \tfrac{1}{2}  a_3\,\varsig) & = \frac{ (1+a_4)(a^2 + l^2-\alpha\,a\,l\,r_H)}
{(a^2 + l^2 - \alpha\,a\,l\,r_H) + \alpha\,a^2\,r_H\,\zeta }. \label{eq:1-Varsigma^2SIMPLE}
\end{align}
Substituting \eqref{eq:HorizonAreaOmega}, \eqref{eq:1-Varsigma^2} and \eqref{eq:1-Varsigma^2SIMPLE} into \eqref{eq:MetricFunctionFact}, we finally obtain
\begin{align}
f_{\text{D}}(\zeta) &=  C\,d_+\, (1 + a_4)^2\,(a^2 + l^2-\alpha\,a\,l\,r_H)^2\, \frac{1-\zeta^2}{ C_0 + C_1 \,\zeta + C_2\,\zeta^2},
\label{eq:MetricFunctionDgeneric}
\end{align}
which is clearly of the same form as the metric function \eqref{eq:MetricFunctionEIH}.

It only remains to compare the coefficients $c_i$ given by \eqref{eq:c0c1c2} with $C_i$ given by \eqref{eq:C0C1C2}.
By substituting \eqref{eq:a3a4explicit}, \eqref{eq:d+-} into \eqref{eq:c0c1c2}, a direct evaluation indeed leads to
\begin{align}\label{Cici}
C_i = c_i\, ( a^2 + l^2 - \alpha\,a\,l\,r_H)^2 , \qquad \hbox{for}\quad i=0,1,2,
\end{align}
provided $q^2$~has a \emph{unique value}
\begin{align*}
q^2 \equiv  \frac{4\pi C}{d_-} \, \frac{C_0-C_2}{( a^2 + l^2 - \alpha\,a\,l\,r_H)^2} .
\end{align*}
\emph{This completes the proof of the equivalence of $f_{\text{D}}(\zeta)$ and $f_{\text{EIH}}(\zeta)$ in a fully generic case}.

Surprisingly, using \eqref{eq:C0C1C2}, the expression for $q^2$ simplifies considerably to
\begin{align}
q^2 & = 4\pi C\,\frac{r_H^2-a^2+l^2}{r_H^2+(a-l)^2}\,
\bigg[ 1- \Big(\,\frac{\alpha\,a^2\,r_H}{ a^2 + l^2 - \alpha\,a\,l\,r_H} \,  \Big)^2     \bigg].
\label{eq:UniqueRelation-generic}
\end{align}
Notice that this formula reduces to \emph{all previous special cases}, namely Eq.~\eqref{eq:UniqueRelation-nontwist} for the non-twisting black holes (at first by setting ${l=0}$ and then ${a=0}$, in which case ${r_H=m}$), Eq.~\eqref{eq:RelationBetweenCharges-nonrot} for the non-rotating black holes (${a=0}$), Eq.~\eqref{eq:UniqueRelation-NoNUT} for accelerating Kerr-Newman black holes (${l=0}$), and Eq.~\eqref{eq:UniqueRelation-alpha=0} for Kerr-Newman-NUT black holes (${\alpha=0}$).

Moreover, using \eqref{eq:Definiceq^2} and  \eqref{eq:HorizonAreaOmega} it  follows that  the \emph{genuine physical electric and magnetic charges} $Q_E$ and $Q_M$ are given by
\begin{align} \label{eq:RelationBetweenCharges-generic}
Q_E^2 + Q_M^2 &= C^2\, \frac{r_H^2+(a+l)^2}{r_H^2+(a-l)^2}\,\frac{r_H^2-a^2+l^2}{(1-\A\,r_H)^2} .
\end{align}
where, according to \eqref{eq:DefA}, the acceleration $\alpha$ of the extremal black hole is only involved through the special combination of the three paremeters, namely
\begin{align*}
\A = \frac{\alpha\,a\,l}{a^2 + l^2}.
\end{align*}

\subsubsection{Accelerating Kerr-NUT black holes: ${e = 0 = g}$}
\label{sec:AcceleratingKerrNUTBlackHoles}

As an important special subcase of the general family of extremal type~D black holes we may finally investigate the \emph{uncharged case}, i.e.,  Kerr-NUT black holes which uniformly accelerate, characterized just by the rotation parameter~$a$, the NUT~parameter~$l$ and the acceleration~$\alpha$. As shown above, these three parameters are combined into the unique quantity $\A$. For extremal black holes of this type, they also fully determine the horizon position $r_H$ and the black hole mass $m$ via \eqref{extremeHorizonGeneralExpl},\eqref{extremeHorizonGeneralExpl2}, or \eqref{extremeHorizonGeneralExplSquare}.

Using \eqref{eq:UniqueRelation-generic}, \eqref{eq:RelationBetweenCharges-generic} we immediately observe that there are
\begin{align} \label{eq:NoCharges}
\hbox{no charges}
  \quad\Leftrightarrow\quad
q^2 = 0
  \quad\Leftrightarrow\quad
Q_E^2 + Q_M^2 = 0
  \quad\Leftrightarrow\quad
r_H^2=a^2-l^2.
\end{align}
Interestingly, it is \emph{equivalent to the condition} ${e=0=g}$. Indeed, substituting ${a^2-l^2 = r_H^2}$ into the expression \eqref{extremeHorizonGeneralExplSquare} we obtain a condition ${(e^2 + g^2)(1+3\A^2r_H^2) = 0}$, which necessarily implies ${e=0=g}$. Thus we have proved that
\begin{align} \label{eq:NoCharges-e=0=g}
\hbox{no charges}
  \quad\Leftrightarrow\quad
r_H^2=a^2-l^2
  \quad\Leftrightarrow\quad
e^2 + g^2 = 0.
\end{align}
Since ${q^2=0}$ implies ${C_0=C_2}$, the corresponding metric function takes an explicit form
\begin{align}
f_{\text{D}}(\zeta) \equiv f_{\text{EIH}}(\zeta) &=  2C a(a+l)\, (1 + a_4)^2\,(a^2 + l^2-\alpha\,a\,l\,r_H)^2\, \frac{1-\zeta^2}{ C_1 \,\zeta + C_2\,(1+\zeta^2)}.
\label{eq:MetricFunctionDgeneric-e=0=g}
\end{align}

\newpage

\section{Summary of the results and concluding remarks}
\label{sec:Summary}

The seminal concept by Ashtekar and Krishnan of isolated horizons (IHs), reviewed in Sec.~\ref{sec:isolated horizons}, provides a prolific mathematical framework for investigation of black holes that are in equilibrium with their neighbourhood. The area of such horizon does not change in time, because there is no flux of matter through it. In order to admit a certain degree of time-dependence in the initial data and allow more general presence of matter, the concept of \emph{weakly isolated horizons} (WIHs) was introduced. Recently, it has been further extended by G\"urlebeck and Scholtz \cite{Scholtz2017} to \emph{almost isolated horizons} (AIHs), which admit more general topologies of the horizon sections, such as axially symmetric compact spatial manifold with deficit angles at poles. In physical terms, such horizon can be pierced by \emph{cosmic strings} or \emph{struts}. This idea was the starting point in our investigation, although we needed a stronger notion of isolation than the AIH.

In their previous work \cite{Lewandowski-Pawlowski}, Lewandowski and Pawlowski considered \emph{extremal isolated horizons} (EIHs), for which they proved uniqueness and local isometry with the Kerr-Newman solution under a strong natural assumption of complete regularity, which excludes the topological defects on the horizon characterized by deficit angles $\delta_+$ and~$\delta_-$. Study of near-horizon geometries of extremal black holes in \cite{KunduriLucietti, KunduriLucietti2009a, KunduriLucietti2009b} provided equivalent results with a possible non-zero cosmological constant.
In this paper, our aim was to extend these results by relaxing the regularity in the above sense, and investigate in detail its physical consequences. As we have already pointed out, topological defects of this kind are inherent feature of many exact spacetimes.

Using the Newman-Penrose (NP) formalism, we systematically studied the \emph{complete} class of axially symmetric EIHs with zero cosmological constant. They are defined geometrically by vanishing surface gravity everywhere on the horizon.
In canonical coordinates ${\zeta, \phi}$, introduced in Eq.~\eqref{CanonicalMetric}, the metric of spatial sections of the \emph{axisymmetric} EIH reads
\begin{align*}
 - R^2\,\Big( \,\frac{\dd \zeta^2}{f(\zeta)} + f(\zeta)\,\dd \phi^2 \,\Big).
\end{align*}
It involves a single dimensionless function $f(\zeta)$ and the radius $R$ defined by the horizon area ${A \equiv 4\pi R^2}$.

In Sec.~\ref{sec:extremal isolated horizons}, we completely integrated the corresponding constraint equations following from the NP formalism (see Appendix~A). Our main result is summarized in Theorem~\ref{th:MetricIH}. In particular, the metric function $f(\zeta)$ must necessarily have the form \eqref{th:MetricFunction}, see also \eqref{eq:MetricFunctionIH}, \eqref{eq:Definiceq^2}, which can be rewritten as
\begin{align*}
  f_{\text{EIH}}(\zeta) &= \frac{2}{\pi}\,  \frac{(2\pi+\delta_-)(2\pi+\delta_+)(1-\zeta^2)}
  {[(\delta_+ + \delta_- + 4\pi)+q^2]
    + 2 (\delta_- - \delta_+)\,\zeta
    + [(\delta_+ + \delta_- + 4\pi)-q^2]\,\zeta^2}.
\end{align*}
The deficit angles $\delta_{\pm}$ are located at the horizon poles corresponding to ${\zeta=\pm1}$, and the dimensionless charge parameter is
\begin{align*}
  q^2 \equiv \frac{(4\pi)^2}{A}\,(Q_E^2 + Q_M^2),
\end{align*}
where $Q_E$ and $Q_M$ are the total electric and magnetic charges, while $A$ is the horizon area. The function $f_{\text{EIH}}$ thus depends on 5 independent parameters $(A, Q_E, Q_M, \delta_-, \delta_+)$.
Without the strings or struts (${\delta_- =0= \delta_+}$) both poles are regular, and our solution simplifies to
\begin{align*}
  f_{\text{EIH}}(\zeta) &= \frac{8\pi\,(1-\zeta^2)}  {(4\pi+q^2)  + (4\pi-q^2)\,\zeta^2},
\end{align*}
recovering the result of \cite{Lewandowski-Pawlowski} for uniqueness of the extremal Kerr-Newman black hole under the assumption of regular spherical topology.

Our second main aim was to compare this locally defined general result with the horizon geometries of a large family of extremal black holes, which are exact spacetimes of algebraic type~D. As summarized in Sec.~\ref{sec:exact type D}, they belong to the Pleba\'{n}ski-Demia\'{n}ski family of \emph{electrovacuum solutions} (without cosmological constant) such that the Maxwell field is double aligned with the gravitational field. We employed the convenient parametrization \eqref{newmetricGP2005} of this family, found by Griffiths and Podolsk\'y \cite{GriffithsPodolsky2006}, which apart from the conicity $C$ (see Sec.~\ref{sec:deficit angles conicity}) includes 6 usual physical parameters $(m, e, g, a, l, \alpha)$ representing the mass, electric and magnetic charges, Kerr-like rotation, NUT parameter, and acceleration of the black hole, respectively. These parameters are here constrained by the horizon extremality condition ${\QQ(r_H)=0=\QQ'(r_H)}$. The \emph{horizon is located at} a specific value of the radial coordinate
\begin{align*}
r_H=\frac{m-\A\,(a^2-l^2+e^2+g^2)+\A^2\, m\,(a^2-l^2)}{1+2\A\, m - 3\A^2(e^2+g^2)},
\end{align*}
where $\A$ is a unique combination of the parameters ${\alpha, a, l}$, namely
\begin{align*}
\A \equiv  \frac{\alpha\,a\,l}{a^2 + l^2},
\end{align*}
see \eqref{extremerootofQ} and \eqref{eq:DefA}--\eqref{extremeHorizonGeneralExplSquare}. Whenever ${\alpha=0}$ or ${a=0}$ or ${l=0}$, it simplifies to ${r_H =  m}$ with ${m^2=a^2-l^2+e^2+g^2}$. Only 5 physical parameters of the extremal black hole are thus independent.

We also derived that the \emph{horizon area} of these extremal black holes has the value
\begin{align*}
A = \frac{ 4\pi C\,[r_H^2+(a+l)^2]\,(a^2 + l^2)^2}{(a^2 + l^2-\alpha\,a\,l\,r_H)^2-\alpha^2a^4\,r_H^2}.
\end{align*}
Contrarily, the area of the two \emph{acceleration} horizons \eqref{accel-horizons} located at ${r_{{\rm a}+}}$ and ${r_{{\rm a}-}}$ is infinite.

Theorem~\ref{th:MetricD} summarizes the main result of Sec.~\ref{sec:exact type D}, that is our derivation of a specific metric function $f_{\text{D}}(\zeta)$, which describes the geometry of the horizon in the complete family of type~D black holes \eqref{newmetricGP2005}. It is given by Eq.~\eqref{eq:MetricFunction},
\begin{align*}
f_{\text{D}}(\zeta) &= \frac{4\pi C^2}{A} \big[r_H^2+(a + l)^2\big]^2
    \frac{\tilde{P}(\zeta)}{\Omega^2(\zeta)\,\rho^2(\zeta)},
\end{align*}
where $\tilde{P}, \Omega, \rho$, introduced in \eqref{eq:PolynomialPtilde}, \eqref{PDmetricGP}, have to be expressed in terms of $\zeta$ via \eqref{eq:ZetaAndVarsigma}.

In the last Sec.~\ref{sec:comparison}, we were able to show that the function $f_{\text{D}}(\zeta)$ \emph{has the same form as} $f_{\text{EIH}}(\zeta)$ for \emph{every} combination of the physical parameters.  Moreover, we found specific relations between the geometrical and physical parameters of the class of type~D black holes, see Sec.~\ref{sec:GeneralBlackHoles}. In particular, we derived an explicit formula for the dimensionless geometrical quantity $q^2$.

The key observation, involved in the comparison of the metric functions $f_{\text{D}}$ and $f_{\text{EIH}}$, was the mutual relation between the constants $a_3$ and $a_4$, which are unique combinations of the physical parameters. In the case of \emph{extremal} black holes, they have to satisfy the relation ${a_3^2=-4\,a_4}$, see \eqref{eq:a3a4extremal}. This implies factorization of the functions $\tilde{P}, \Omega, \rho$ and further simplifications of the coefficients, see \eqref{eq:extremalP}--\eqref{eq:d+-}. Putting it together, this yields the metric function in the form
\begin{align*}
f_{\text{D}}(\zeta) = C\,[r_H^2+(a+l)^2](a^2 + l^2-\alpha\,a\,l\,r_H)^2 \,  \frac{(1 + a_4)^2 \, (1-\zeta^2)}{ C_0 + C_1 \,\zeta + C_2\,\zeta^2},
\end{align*}
where $a_4$ is given by \eqref{eq:a3a4explicit} and the constants $C_i$ by \eqref{eq:C0C1C2}. This function is equivalent to $f_{\text{EIH}}(\zeta)$, expressed via the physical parameters \eqref{eq:MetricFunctionEIH}, provided we choose the dimensionless charge parameter $q^2$ as
\begin{align*}
q^2  = 4\pi C\,\frac{r_H^2-a^2+l^2}{r_H^2+(a-l)^2}\,
\bigg[ 1- \Big(\,\frac{\alpha\,a^2\,r_H}{ a^2 + l^2 - \alpha\,a\,l\,r_H} \,  \Big)^2  \bigg],
\end{align*}
see \eqref{eq:UniqueRelation-generic}.
The reason why we had to find such $q^2$ is that the physical parameters of $f_{\text{D}}$ have well-understood meaning only in particular special cases, as we have demonstrated in the case of electric and magnetic charges. With this choice, the \emph{total electric and magnetic charges} of such extremal black holes are
\begin{align*}
Q_E^2 + Q_M^2 = C^2\, \frac{r_H^2+(a+l)^2}{r_H^2+(a-l)^2}\,\frac{r_H^2-a^2+l^2}{(1-\A\,r_H)^2}.
\end{align*}
It seems that only in the case of non-rotating black holes (${a=0}$) or black holes without the NUT parameter (${l=0}$) this formula reduces to a simple relation
\begin{align*}
Q_E^2 + Q_M^2 = C^2\,(r_H^2-a^2+l^2)=C^2\,(e^2+g^2),
\end{align*}
and the genuine electromagnetic charges $Q_E, Q_M$ are then proportional to the electromagnetic parameters in the type~D metric just via the conicity $C$,
\begin{align*}
Q_E = C\,e,\qquad
Q_M = C\,g.
\end{align*}

We thus clarified how different physical parameters influence the geometry of extremal black hole horizon. From our discussion of the individual subcases it also follows that \emph{different}  extremal black holes might have \emph{the same} horizon structure represented by $f_{\text{EIH}}$. For instance, extremal Reissner-Nordstr\"om black hole has the horizon geometry isometric to the one of extremal charged NUT solution, see Sec.~\ref{sec:NonRotatingBlackHoles}. Hence, we can not distinguish between the spacetimes just from the knowledge of the horizon geometry.

The general expressions for extremal black holes considerably simplify in various interesting subclasses, such as for non-twisting black holes investigated in Sec.~\ref{sec:NonTwistingBlackHoles} (these are accelerating extremely charged Reissner-Nordstr\"om black holes, i.e. the C-metric with ${e^2=m^2}$), extremely charged NUT black holes (Sec.~\ref{sec:NonRotatingBlackHoles}), extremal accelerating Kerr-Newman black holes (Sec.~\ref{sec:AccKerrNew}),  non-accelerating Kerr-Newman-NUT black holes (Sec.~\ref{sec:KerrNewmanNUTBlackHoles}), or accelerating Kerr-NUT black holes (Sec.~\ref{sec:AcceleratingKerrNUTBlackHoles}).

As a natural extension of the present work, a non-zero value of cosmological constant~$\Lambda$ could be considered. Previously, analogous solutions of this kind were obtained, see the review \cite{KunduriLucietti} and the recent article \cite{BukLewandowski}. These, however, did not analyze the most general solution with clearly identified parameters and their physical interpretation. Another interesting question is whether the axial symmetry of the extremal isolated horizon could be relaxed. It was already shown that extremality implies axial symmetry in asymptotically flat spacetimes \cite{Amsel2009, HollandsIshibashi, ChruscielCosta}.

Finally, we would also like to point the reader to  recently published  works \cite{Lewandowski-Szereszewski, Lewandowski-Ossowski, Lewandowski-OssowskiII, Dobkowski-Lewandowski-Pawlowski-2018} in which a different (complementary) approach to regularity was addopted.


\section*{Acknowledgements}

We thank both referees for their comments and suggestions.
This paper has been supported by the Czech Science Foundation Grant No.~19-01850S.


\appendix

\section{Newman-Penrose formalism}
\label{chap:NPFormalism}

For the convenience of the reader, we offer here a brief summary of definitions and equations of the Newman-Penrose formalism.
\subsection{Gravitational field}

\noindent Directional derivatives:
\begin{align} \label{NP:Derivatives}
D&=\ell^a\,\nabla_a, &
\Delta &= n^a \,\nabla_a, &
\delta &= m^a \,\nabla_a, &
\bar{\delta} &= \bar{m}^a \,\nabla_a.
\end{align}

\noindent Decomposition of the covariant derivative:
\begin{align}\label{NP:NablaDecomposition}
 \nabla_a = {g_a}^b\, \nabla_b  = \ell_a \,\Delta + n_a\,D - m_a\,\bar{\delta} -
\bar{m}_a\,\delta.
\end{align}

\noindent Spin coefficients:
\begin{subequations}
\begin{align} \label{NP:SpinCoefficients}
 \kappa &= m^a D \ell_a , &
  \tau &= m^a \Delta \ell_a ,  \nonumber \\
\sigma &= m^a \delta \ell_a , &
   \rho& = m^a \bar{\delta} \ell_a , \nonumber \\
\pinp &= n^a D \bar{m}_a , &
  \nu &= n^a \Delta \bar{m}_a , \nonumber \\
 \lambda &= n^a \bar{\delta} \bar{m}_a ,
&
  \mu &= n^a \delta \bar{m}_a , \\
 \eps &= \textstyle \frac{1}{2}\left(
n^a D \ell_a - \bar{m}^a D m_a\right) , &
\beta &= \textstyle \frac{1}{2}\left(
n^a \delta \ell_a - \bar{m}^a \delta m_a\right) , \nonumber \\
 \gamma &= \textstyle \frac{1}{2}\left(
n^a \Delta \ell_a - \bar{m}^a \Delta m_a\right) , &
  \alpha &= \textstyle \frac{1}{2}\left(
n^a \bar{\delta} \ell_a - \bar{m}^a \bar{\delta} m_a\right) , \nonumber
\end{align}
\end{subequations}

\noindent The operators \eqref{NP:Derivatives} acting on a scalar function obey this commutation
relations:
\begin{subequations} \label{NP:Commutators}
\begin{align}
D\delta - \delta D  &=  (\pinpb-\bar{\alpha}-\beta)D - \kappa
\Delta + (\bar{\rho}-\bar{\varepsilon}+\varepsilon)\delta + \sigma
\bar{\delta},\label{np:CR:Ddelta}\\
\Delta D  - D \Delta  &=
(\gamma+\bar{\gamma})D +(\varepsilon + \bar{\varepsilon})\Delta -
(\bar{\tau}+\pinp)\delta -
(\tau+\pinpb)\bar{\delta},\label{np:CR:DeltaD} \\
\Delta \delta
- \delta \Delta &=  \bar{\nu}D + (\bar{\alpha}+\beta -
\tau)\Delta + (\gamma-\bar{\gamma}-\mu)\delta - \bar{\lambda}
\bar{\delta},\label{np:CR:Deltadelta}\\
\delta\bar{\delta} - \bar{\delta} \delta
&=  (\mu-\bar{\mu})D + (\rho-\bar{\rho})\Delta +
(\bar{\alpha}-\beta)\bar{\delta} -
(\alpha-\bar{\beta})\delta.\label{np:CR:deltadeltabar}
\end{align}
\end{subequations}

\noindent Transport equations:
\begin{subequations} \label{NP:TransportEquations}
\begin{align}
 D\ell^a &= \left( \eps+\bar{\eps} \right) \ell^a - \bar{\kappa}\,m^a -
\kappa\,\bar{m}^a, \label{np:transport eqs-D la}\\
\Delta \ell^a &= \left( \gamma+\bar{\gamma} \right)\ell^a-\bar{\tau}\,m^a -
\tau\,\bar{m}^a, \label{np:transport eqs-Delta la}\\
\delta \ell^a &= \left( \bar{\alpha}+\beta \right)\ell^a-\bar{\rho}\,m^a -
\sigma\,\bar{m}^a, \label{np:transport eqs-delta la}\\
Dn^a &= -\left( \eps+\bar{\eps} \right)n^a +\pinp\,m^a
+\pinpb\,\bar{m}^a,\label{np:transport eqs-D na}\\
\Delta n^a &= -\left( \gamma+\bar{\gamma} \right)n^a +\nu\,m^a +
\bar{\nu}\,\bar{m}^a, \label{np:transport eqs-Delta na}\\
\delta n^a &= -(\bar{\alpha}+\beta)n^a+\mu\,m^a + \bar{\lambda}\,\bar{m}^a,
\label{np:transport eqs-delta na}\\
D m^a &= \pinpb\,\ell^a - \kappa\,n^a + \left( \eps-\bar{\eps} \right)m^a,
\label{np:transport eqs-D ma} \\
\Delta m^a &= \bar{\nu}\,\ell^a -\tau\,n^a + \left( \gamma-\bar{\gamma}
\right)m^a, \label{np:transport eqs-Delta ma}\\
\delta m^a &= \bar{\lambda}\,\ell^a-\sigma\,n^a+\left( \beta-\bar{\alpha}
\right)m^a, \label{np:transport eqs-delta ma} \\
\bar{\delta}m^a &= \bar{\mu}\,\ell^a -\rho\,n^a+\left( \alpha-\bar{\beta}
\right)m^a.\label{np:transport eqs-deltabar ma}
\end{align}
\label{np:transport eqs}
\end{subequations}

\noindent The Riemann tensor can be decomposed into the Weyl tensor $C_{abcd}$, the trace-free part of the Ricci tensor, and the scalar $\Lambda$ related to the scalar curvature $R$ by
\begin{align}
\Lambda &=  \frac{1}{24}R.
\end{align}

\noindent The five (complex) tetrad components of the Weyl spinor are
\begin{subequations} \label{NP:WeylScalars}
\begin{align}
\Psi_0 &= C_{abcd}\,l^a m^b l^c m^d ,\label{psicomps2}\\
\Psi_1 &= C_{abcd}\,l^a n^b l^c m^d ,\\
\Psi_2 &= C_{abcd}\,l^a m^b\bar{m}^c n^d ,\\
\Psi_3 &= C_{abcd}\,l^a n^b\bar{m}^c n^d ,\\
\Psi_4 &= C_{abcd}\,\bar{m}^a n^b\bar{m}^cn^d .
\end{align}
\end{subequations}

\noindent The traceless Ricci tensor (equal to $R_{ab}$ when ${\Lambda=0}$) has the following components, from which are 3 real and 3 are complex:
\begin{subequations} \label{NP:RicciComponents}
\begin{align}
\Phi_{00} &= - \textstyle \frac{1}{2} R_{ab}\, l^a l^b  ,
\label{RicciComps}\\
\Phi_{01} &= - \textstyle \frac{1}{2} R_{ab}\, l^a m^b ,\\
\Phi_{02} &= - \textstyle \frac{1}{2} R_{ab}\, m^a m^b  ,\\
\Phi_{11} &= - \textstyle \frac{1}{4} R_{ab} \left(l^a n^b + m^a \bar{m}^b\right)  ,\\
\Phi_{12} &= - \textstyle \frac{1}{2} R_{ab}\, n^a m^b  ,\\
\Phi_{22} &= - \textstyle \frac{1}{2} R_{ab}\, n^a n^b  .
\end{align}
\end{subequations}
The three remaining components can be obtained via the
symmetry $\Phi_{ij}=\bar{\Phi}_{ji}$.

\noindent The Ricci identities read:
\begin{subequations} \label{NP:RicciIdentities}
\begin{align}
  D \rho - \bar{\delta} \kappa &=\rho ^2+\left(\epsilon +\bar{\epsilon
   }\right) \rho -\kappa  \left(3 \alpha +\bar{\beta }-\pinp \right)-\tau
   \bar{\kappa }+\sigma  \bar{\sigma }+\Phi_{00},\label{np:RI:Drho}\\
  D\sigma-\delta\kappa &= (\rho+\bar{\rho}+3\eps-\bar{\eps})\sigma -
(\tau-\pinpb+\bar{\alpha}+3\beta)\kappa+\Psi_0,\label{np:RI:Dsigma}\\
  D\tau-\Delta\kappa &=
\rho(\tau+\pinpb)+\sigma(\bar{\tau}+\pinp)+(\eps-\bar{\eps})\tau
-(3\gamma+\bar{\gamma})\kappa+\Psi_1+\Phi_{01},\label{np:RI:Dtau}\\
  D\alpha-\bar{\delta}\eps &= (\rho
+\bar{\eps}-2\eps)\alpha+\beta\bar{\sigma}-\bar{\beta}\eps - \kappa \lambda -
\bar{\kappa}\gamma + (\eps+\rho)\pinp + \Phi_{10},\label{np:RI:Dalpha}\\
  D\beta-\delta\eps &= (\alpha+\pinp)\sigma +
(\bar{\rho}-\bar{\eps})\beta-(\mu+\gamma)\kappa-(\bar{\alpha}-\pinpb)\eps +
\Psi_1,\label{np:RI:Dbeta}\\
  D\gamma-\Delta\eps &= (\tau+\pinpb)\alpha + (\bar{\tau}+\pinp)\beta -
(\eps+\bar{\eps})\gamma - (\gamma + \bar{\gamma})\eps \nonumber \\
& \quad + \tau \pinp - \nu
\kappa
 + \Psi_2 - \Lambda + \Phi_{11},\label{np:RI:Dgamma}\\
  D\lambda-\bar{\delta}\pinp &= (\rho - 3\eps+\bar{\eps})\lambda +
\bar{\sigma}\mu + (\pinp+\alpha-\bar{\beta})\pinp -
\nu\bar{\kappa}+\Phi_{20},\label{np:RI:Dlambda}\\
  D\mu-\delta\pinp &= (\bar{\rho}-\eps-\bar{\eps})\mu+\sigma\lambda+
(\pinpb-\bar{\alpha}+\beta)\pinp - \nu \kappa + \Psi_2 + 2
\Lambda,\label{np:RI:Dmu}\\
  D\nu-\Delta\pinp &=
(\pinp+\bar{\tau})\mu+(\pinpb+\tau)\lambda+(\gamma-\bar{\gamma})\pinp -
(3\eps+\bar{\eps})\nu+\Psi_3+\Phi_{21},\label{np:RI:Dnu}\\
  \Delta\lambda-\bar{\delta}\nu &=
-(\mu+\bar{\mu}+3\gamma-\bar{\gamma})\lambda+(3\alpha+\bar{\beta}+\pinp-\bar{\tau}
)\nu-\Psi_4,\label{np:RI:Deltalambda}\\
  \Delta\mu-\delta\nu &=
-(\mu+\gamma+\bar{\gamma})\mu-\lambda\bar{\lambda}+\bar{\nu}\pinp+(\bar{\alpha}
+3\beta-\tau)\nu-\Phi_{22},\label{np:RI:Deltamu}\\
  \Delta\beta-\delta\gamma &= (\bar{\alpha}+\beta-\tau)\gamma - \mu \tau +
\sigma \nu + \eps \bar{\nu} + (\gamma-\bar{\gamma}-\mu)\beta -
\alpha\bar{\lambda}-\Phi_{12},\label{np:RI:Deltabeta}\\
  \Delta\sigma-\delta\tau& = -(\mu-3\gamma+\bar{\gamma})\sigma -
\bar{\lambda}\rho - (\tau + \beta - \bar{\alpha})\tau + \kappa
\bar{\nu}-\Phi_{02},\label{np:RI:Deltasigma}\\
  \Delta\rho-\bar{\delta}\tau &= (\gamma+\bar{\gamma}-\bar{\mu})\rho - \sigma
\lambda + (\bar{\beta}-\alpha-\bar{\tau})\tau + \nu \kappa - \Psi_2 - 2
\Lambda,\label{np:RI:Deltarho}\\
  \Delta\alpha-\bar{\delta}\gamma &= (\rho+\eps)\nu - (\tau+\beta)\lambda +
(\bar{\gamma}-\bar{\mu})\alpha + (\bar{\beta}-\bar{\tau})\gamma -
\Psi_3,\label{np:RI:Deltaalpha}\\
  \delta\rho-\bar{\delta}\sigma &= (\bar{\alpha}+\beta)\rho -
(3\alpha-\bar{\beta})\sigma+(\rho-\bar{\rho})\tau+(\mu-\bar{\mu})\kappa -\Psi_1
+ \Phi_{01},\label{np:RI:deltarho}\\
  \delta\alpha-\bar{\delta}\beta &= \mu\rho-\lambda\sigma +
\alpha\bar{\alpha}+\beta\bar{\beta}-2\alpha\beta + (\rho-\bar{\rho})\gamma +
(\mu-\bar{\mu})\eps  \nonumber \\
& \quad -  \Psi_2 + \Lambda + \Phi_{11},\label{np:RI:deltaalpha}\\
  \delta\lambda-\bar{\delta}\mu &= (\rho-\bar{\rho})\nu + (\mu-\bar{\mu})\pinp
+ (\alpha+\bar{\beta})\mu+(\bar{\alpha}-3\beta)\lambda-\Psi_3 +
\Phi_{21}.\label{np:RI:deltalambda}
\end{align}
\end{subequations}

\noindent The Bianchi identities in the NP formalism are:
\begin{subequations}
\begin{multline}
 D\Psi_1-\bar{\delta}\Psi_0-D\Phi_{01}+\delta\Phi_{00} = (\pinp - 4 \alpha)
\Psi_0+2(2\rho+\varepsilon)\Psi_1-3\kappa\Psi_2+2\kappa\Phi_{11}  -
(\pinpb-2\bar{\alpha}-2\beta)\Phi_{00}
\\-2\sigma\Phi_{10}-
2(\bar{\rho}+\varepsilon)\Phi_{01}+\bar{\kappa}\Phi_{02},\label{np:BI:DPsi1}
\end{multline}
\begin{multline}
 D\Psi_2-\bar{\delta}\Psi_1+\Delta\Phi_{00}-\bar{\delta}\Phi_{01}
+2D\Lambda =-\lambda\Psi_0 + 2 (\pinp-\alpha)\Psi_1+3\rho
\Psi_2-2\kappa\Psi_3+2\rho\Phi_{11}+\bar{\sigma}\Phi_{02}
\\
+ (2\gamma+2\bar{\gamma}-\bar{\mu})\Phi_{
00}-2(\alpha+\bar{\tau})\Phi_{01}-2\tau\Phi_{10},\label{np:BI:DPsi2}
\end{multline}
\begin{multline}
 D\Psi_3-\bar{\delta}\Psi_2-D\Phi_{21}+\delta\Phi_{20}-2\bar{\delta}\Lambda
= -2\lambda \Psi_1+3\pinp\Psi_2 + 2
(\rho-\varepsilon)\Psi_3-\kappa\Psi_4+2\mu\Phi_{10}- 2\pinp\Phi_{11}\\
-(2\beta+\pinpb-2\bar{\alpha})\Phi_{20}-2(\bar{
\rho}-\varepsilon)\Phi_{21}+\bar{\kappa}\Phi_{22},\label{np:BI:DPsi3}
\end{multline}
\begin{multline}
D\Psi_4-\bar{\delta}\Psi_3+\Delta\Phi_{20}-\bar{\delta}\Phi_{21}
=-3\lambda\Psi_2
+2(\alpha+2\pinp)\Psi_3+(\rho-4\varepsilon)\Psi_4+2\nu\Phi_{10}-2\lambda\Phi_{11}
\\
- (2\gamma-2\bar{\gamma}+\bar{\mu})\Phi_{20}-2(\bar{\tau}
-\alpha)\Phi_{21}+\bar{\sigma}\Phi_{22},\label{np:BI:DPsi4}
\end{multline}
\begin{multline}
\Delta\Psi_0-\delta\Psi_1+D\Phi_{02}-\delta\Phi_{01}
=(4\gamma-\mu)\Psi_0-2(2\tau+\beta)\Psi_1+3\sigma\Psi_2\\
+(\bar{\rho}+2\varepsilon-2\bar{\varepsilon})\Phi_{02}+ 2\sigma\Phi_{11}
-2\kappa\Phi_{12}-\bar{\lambda}\Phi_{00}+2(\pinpb-\beta)\Phi_{01},
\label{np:BI:DeltaPsi0}
\end{multline}
\begin{multline}
\Delta\Psi_1-\delta\Psi_2-\Delta\Phi_{01}+\bar{\delta}\Phi_{02}-2\delta\Lambda
=\nu\Psi_0+2(\gamma-\mu)\Psi_1-3\tau\Psi_2+2\sigma\Psi_3\\
-\bar{\nu}\Phi_{00}+
2(\bar{\mu}-\gamma)\Phi_{01}+(2\alpha+\bar{\tau}-2\bar{\beta})\Phi_{02}
+2\tau\Phi_{11}-2\rho\Phi_{12},
\label{np:BI:DeltaPsi1}
\end{multline}
\begin{multline}
\Delta\Psi_2-\delta\Psi_3+D\Phi_{22}-\delta\Phi_{21}
+2\Delta\Lambda=2\nu\Psi_1-3\mu\Psi_2+2(\beta-\tau)\Psi_3+\sigma\Psi_4\\
-2\mu\Phi_{11}-\bar{\lambda}\Phi_{20}+ 2\pinp\Phi_{12}+2(\beta+\pinpb)\Phi_{21
}+(\bar{\rho}-2\varepsilon-2\bar{\varepsilon})\Phi_{22},
\label{np:BI:DeltaPsi2}
\end{multline}
\begin{multline}
\Delta\Psi_3-\delta\Psi_4-\Delta\Phi_{21}+\bar{\delta}\Phi_{22}
=3\nu\Psi_2-2(\gamma+2\mu)\Psi_3+(4\beta-\tau)\Psi_4-2\nu\Phi_{11}\\
-\bar{\nu}\Phi_{20}+2\lambda\Phi_{12}+2(\gamma+\bar{\mu})\Phi_{21}+(\bar{\tau}
-2\bar{\beta}-2\alpha)\Phi_{22},
\label{np:BI:DeltaPsi3}
\end{multline}
\begin{multline}
 D\Phi_{11}-\delta\Phi_{10}+\Delta\Phi_{00}-\bar{\delta}\Phi_{01}+3D\Lambda =
(2\gamma+2\bar{\gamma}-\mu-\bar{\mu})\Phi_{00}+(\pinp-2\alpha-2\bar{\tau})\Phi_{01
}\\
+ (\pinpb-2\bar{\alpha}-2\tau)\Phi_{10}+2(\rho+\bar{\rho})\Phi_{11}+\bar{
\sigma}\Phi_{02}
+\sigma\Phi_{20}-\bar{\kappa}\Phi_{12}-\kappa\Phi_{21},
\label{np:BI:DPhi11}
\end{multline}
\begin{multline}
D\Phi_{12}-\delta\Phi_{11}+\Delta\Phi_{01}-\bar{\delta}\Phi_{02}+3\delta\Lambda
= (2\gamma-\mu-2\bar{\mu})\Phi_{01}+\bar{\nu}\Phi_{00}-\bar{\lambda}\Phi_{10}\\
+ 2(\pinpb-\tau)\Phi_{11}+(\pinp+2\bar{\beta}-2\alpha-\bar{\tau}
)\Phi_{02}
+(2\rho+\bar{\rho}-2\bar{\varepsilon})\Phi_{12}+\sigma\Phi_{21}-\kappa\Phi_{22},
\label{np:BI:DPhi12}
\end{multline}
\begin{multline}
 D\Phi_{22}-\delta\Phi_{21}+\Delta\Phi_{11}-\bar{\delta}\Phi_{12}+3\Delta\Lambda
= \nu
\Phi_{01}+\bar{\nu}\Phi_{10}-2(\mu+\bar{\mu})\Phi_{11}-\lambda\Phi_{02}-\bar{
\lambda}\Phi_{20}\\
+ (2\pinp-\bar{\tau}+2\bar{\beta})\Phi_{12}+(2\beta-\tau+2\pinpb
)\Phi_{21}
+ (\rho+\bar{\rho}-2\varepsilon-2\bar{\varepsilon})\Phi_{22}.
\label{np:BI:DPhi22}
\end{multline}

\end{subequations}

\subsection{Electromagnetic field} \label{NP:sec:ElmagField}

Electromagnetic field is described by the antisymmetric electromagnetic tensor $F_{ab}$. Its projections onto a NP tetrad are
\begin{subequations}
\begin{align}
\phi_0 &\equiv  F_{ab}\, l^a m^b  ,  \\
\phi_1 &\equiv  \textstyle \frac{1}{2} F_{ab} \left(l^a n^b  +  m^a
\bar{m}^b\right)  , \\
\phi_2 &\equiv  F_{ab} \,\bar{m}^a n^b  ,
\label{np:EM:phi components}
\end{align}
\end{subequations}
The source free Maxwell equations in the NP formalism read
\begin{subequations}
\begin{align}
D\phi_1 - \bar{\delta}\phi_0 &=  (\pinp -
2\alpha)\phi_0 + 2\rho\phi_1 - \kappa \phi_2, \label{NP:Dphi1}\\
D\phi_2 - \bar{\delta}\phi_1 &=  -\lambda \phi_0  +
2\pinp\phi_1 + (\rho-2\varepsilon)\phi_2,\label{NP:Dphi2}\\
\Delta\phi_0 - \delta\phi_1 &=  (2\gamma
-\mu)\phi_0 - 2\tau\phi_1 + \sigma\phi_2,\label{NP:Deltaphi0}\\
\Delta\phi_1 - \delta\phi_2 &=  \nu
\phi_0 - 2\mu\phi_1 + (2\beta-\tau)\phi_2.\label{NP:Deltaphi1}
\end{align}
\end{subequations}
The Ricci tensor in electrovacuum spacetimes is given by
\begin{align}
 \Phi_{mn} &= 2\,\phi_m\,\bar{\phi}_n, \quad \Lambda = 0,
 \label{np:EM:Einstein eqs}
\end{align}
due to the Einstein equations  in the case of vanishing cosmological constant.

\subsection{Spin transformation}
\label{sec:spin-transf}

The spin transformation is defined as rotation in the plane spanned by $m^a, \conj{m}^a$,
\begin{align} \label{NP:SpinTransformation}
m^a \mapsto e^{\ii\,\chi}\,m^a, \quad  \bar{m}^a &\mapsto e^{-\ii\,\chi}\,\bar{m}^a,
\end{align}
where $\chi$ is arbitrary real function. The scalar quantity $\eta$ is said to
have the \emph{spin weight~s} provided
\begin{align*}
\eta' = e^{\ii\,s\,\chi}\,\eta.
\end{align*}
under transformation \eqref{NP:SpinTransformation}. The NP operators $\delta$ and $\bar{\delta}$ contain vectors $m^a, \conj{m}^a$ and do not preserve the
spin weight. If $\eta$ has spin weight $s$, we obtain
\begin{align*}
\delta'\eta'  = e^{\ii (s+1)\chi} \left(\delta\eta +
\ii \, s \,\eta \, \delta \chi \right).
\end{align*}
The prefactor suggests that $\delta \eta$ could have the weight ${s+1}$, but there is an
inhomogeneous term proportional to $\delta\chi$. However, this term can be
eliminated defining a new operator $\eth$ by
\begin{align} \label{NP:Eth}
\eth \eta & \equiv \delta\eta + s(\bar{\alpha}-\beta)\,\eta = \delta\eta + s\,\bar{a}\,\eta,
\end{align}
where
\begin{align} \label{NP:defa}
 a & \equiv \alpha-\bar{\beta},
\end{align}
which transforms homogeneously:
\begin{align*}
 \eth\eta \mapsto e^{\ii (s+1) \chi} \, \eth\eta.
\end{align*}
So if $\eta$ has the spin weight $s$ then $\eth \eta$ has the spin weight $s+1$. Therefore, $\eth$ acts as a spin-raising operator. Analogously we define
\begin{align} \label{NP:Ethbar}
\bar{\eth} \eta &= \bar{\delta}\eta - s(\alpha-\bar{\beta})\,\eta = \delta\eta - s\, a\, \eta.
\end{align}
The spin weights of the Newman-Penrose quantities are summarized in Tab.~\ref{tab:spin weights}.

\begin{table}[h]
 \begin{center}
\begin{align}
\begin{array}{|ccccc|}
\hline
-2 & -1    & 0      & 1      & 2\\
\hline
 \lambda  &  \nu  & \rho   & \kappa & \sigma \\
   & \pinp   & \mu    & \tau & \\
   & \phi_2& \phi_1 & \phi_0 & \\
 \Psi_4 & \Psi_3 & \Psi_2 & \Psi_1 & \Psi_0 \\
\hline
\end{array}
\end{align}
\end{center}
\caption{Spin weights of the Newman-Penrose scalars.}
\label{tab:spin weights}
\end{table}

\subsection{Lorentz transformations}

Let us recall the transformation properties of the  NP  spin coefficients under a particular Lorentz transformation of the null tetrad, namely the null rotation about $l^a$:
\begin{align*}
l'^a = l^a, \quad m'^a = m^a + \conj{c}\, l^a, \quad n'^a = n^a + c\, m^a +  \conj{c}\,\conj{m}^a + c \conj{c}\, l^a,
\end{align*}
for other transformations see \cite{Stephani} and appendix B in \cite{Stewart-1993}.
They transform as
\begin{align}
\begin{split} \label{NP:NullRotationSpinCoefficients}
\kappa' &= \kappa ,  \\
\eps' &= \eps + c \kappa ,  \\
\sigma' &= \sigma + \conj{c} \kappa ,  \\
\rho' &= \rho + c \kappa ,  \\
\tau' &= \tau + c \sigma + \conj{c} \rho + c \conj{c} \kappa ,  \\
\alpha' &= c \alpha \eps + c \rho + c^2 \kappa ,  \\
\beta' &= \beta + c \sigma + \conj{c}\eps +  c \conj{c} \kappa ,  \\
\pinp' &= \pinp + 2c \eps + c^2 \kappa + Dc ,  \\
\gamma' &= \gamma + \conj{c} \alpha + c(\tau + \beta) + c\conj{c}(\rho + \eps) + c^2\sigma + c^2 \conj{c} \kappa ,  \\
\lambda' &= \lambda + c\pinp + 2c\alpha + c^2(\rho + 2\eps) + c^3\kappa + cDc + \conj{\delta} c ,  \\
\mu' &= \mu + 2c\beta + \conj{c}\pinp + c^2 \sigma + 2c \conj{c} \eps  + c^2 \conj{c} \kappa + \conj{c} D c + \delta c ,  \\
\nu' &= \nu + c(2\gamma + \mu) + \conj{c} \lambda + c^2(\tau + 2\beta) + c\conj{c} (\pinp + 2\alpha)   \\
& \quad + c^3 \sigma + c^2 \conj{c} (\rho + 2\eps) + c^3\conj{c} \kappa
 + \Delta c + c\delta c + \conj{c} \conj{\delta} c + c\conj{c}Dc.
\end{split}
\end{align}
For transformations of the Weyl, Riemann and Ricci tensor components see \cite{Stephani,Stewart-1993}.


\addcontentsline{toc}{section}{References}
\begin{thebibliography}{10}

\bibitem{Kerr1963}
R.~P. Kerr,  Gravitational field of a spinning mass as an example of
  algebraically special metrics, {\em Phys. Rev. Lett.} {\bf 11}, 237 (1963).

\bibitem{Newman1965}
E.~T. Newman, E.~Couch, K.~Chinnapared, A.~Exton, A.~Prakash, and R.~Torrence,
   Metric of a rotating, charged mass,  {\em J. Math. Phys.} {\bf 6},  918 (1965).

\bibitem{Reis}
R.~Reis, M.~T. Reynolds, J.~M. Miller, and D.~J. Walton,  Reflection from the
  strong gravity regime in a lensed quasar at redshift $z = 0.658$,
  {\em  Nature} {\bf 24},  207 (2014).

\bibitem{Blandford-Znajek}
R.~D. Blandford and R.~L. Znajek, Electromagnetic extraction of energy from
  Kerr black holes,  {\em Monthly Notices R. Astron. Soc.}
  {\bf 179},  433 (1977).

\bibitem{Gurlebeck2015}
N.~G\"urlebeck,  No-hair theorem for black holes in astrophysical
  environments,  {\em Phys. Rev. Lett.} {\bf 114}, 151102 (2015).

\bibitem{Psaltis2016}
D.~Psaltis, N.~Wex, and M.~Kramer,  A quantitative test of the no-hair theorem
  with Sgr A* using stars, pulsars and the event horizon telescope,
  {\em  Astrophys. J.} {\bf 818}, 121 (2016).

\bibitem{Ashtekar2004-review}
A.~Ashtekar and B.~Krishnan,  Isolated and dynamical horizons and their
  applications,  {\em Living Rev. Relativ.} {\bf 7}, 10 (2004).

\bibitem{Scholtz2017}
N.~G\"urlebeck and M.~Scholtz,  Meissner effect for weakly isolated
  horizons,  {\em Phys. Rev. D} {\bf 95}, 064010 (2017).

\bibitem{Scholtz2018}
N.~G\"urlebeck and M.~Scholtz,  Meissner effect for axially symmetric charged
  black holes,  {\em Phys. Rev. D} {\bf 97}, 084042 (2018).

\bibitem{Griffiths2009}
J.~B. Griffiths and J.~Podolsk\'y,
{\em Exact Space-Times in Einstein's General Relativity}
(Cambridge University Press, Cambridge, England, 2009).

\bibitem{Bardeen1973}
J.~M.~Bardeen, B.~Carter, and S.~W.~Hawking, The four laws of black hole mechanics,
{\em Commun. Math. Phys.} {\bf 31}, 161 (1973).

\bibitem{KunduriLucietti}
H.~K.~Kunduri and J.~Lucietti, Classification of near-horizon geometries of extremal black holes,
 {\em Living Rev. Relativ.} {\bf 16}, 8 (2013).

\bibitem{PodolskySvarc2013a}
J.~Podolsk\'y and R.~\v{S}varc,  Explicit algebraic classification of
  Kundt geometries in any dimension, {\em Class. Quantum Grav.}
  {\bf 30}, 125007 (2013).

\bibitem{PodolskySvarc2013b}
J.~Podolsk{\'{y}} and R.~{\v{S}}varc,  Physical interpretation of Kundt
  spacetimes using geodesic deviation {\em Class. Quantum Grav.}
  {\bf 30}, 205016 (2013).

\bibitem{Lewandowski-Pawlowski}
J.~Lewandowski and T.~Pawlowski,  Extremal isolated horizons: a local
  uniqueness theorem,  {\em Class. Quantum Grav.} {\bf 20}, 587 (2003).

\bibitem{KunduriLucietti2009a}
H.~K.~Kunduri and J.~Lucietti, A classification of near-horizon geometries of extremal vacuum black holes,
{\em J. Math. Phys.} {\bf 50}, 082502 (2009).

\bibitem{KunduriLucietti2009b}
H.~K.~Kunduri and J.~Lucietti, Uniqueness of near-horizon geometries of rotating extremal $AdS_4$ black holes, {\em Class. Quantum Grav.} {\bf 26}, 055019 (2009).

\bibitem{LiLucietti2013}
C.~Li and J.~Lucietti, Uniqueness of extreme horizons in Einstein-Yang-Mills theory, {\em Class. Quantum Grav.} {\bf 30}, 095017 (2013).

\bibitem{Hajicek}
P.~H\'{a}j\'{i}\v{c}ek, Three remarks on axisymmetric stationary horizons, {\em Commun. Math. Phys.} {\bf 36}, 305 (1974).

\bibitem{Amsel2009}
A.~J.~Amsel,  G.~T.~Horowitz, D.~Marolf, and M.~M.Roberts, Uniqueness of extremal Kerr and Kerr-Newman black holes, {\em Phys. Rev. D} {\bf 81}, 024033 (2010).

\bibitem{Stewart-1993}
J.~Stewart, {\em Advanced General Relativity}
(Cambridge University Press, Cambridge, England, 1993).

\bibitem{Stephani}
H.~Stephani, D.~Kramer, M.~MacCallum, C.~Hoenselaers, and E.~Herlt,
{\em Exact Solutions of Einstein's Field Equations}
(Cambridge University Press, Cambridge, England, 2003).

\bibitem{Krishnan-2012}
B.~Krishnan, The spacetime in the neighborhood of a general isolated black
  hole,  {\em Class. Quantum Grav.} {\bf 29}, 205006 (2012).

\bibitem{Ashtekar2004-multipolemoments}
A.~Ashtekar, J.~Engle, T.~Pawlowski, and C.~V.~D. Broeck,  Multipole moments
  of isolated horizons,  {\em Class. Quantum Grav.} {\bf 21},  2549 (2004).

\bibitem{Goldberg1967}
J.~N. Goldberg, A.~J. MacFarlane, E.~T. Newman, F.~Rohrlich, and E.~C.~G.
  Sudarshan,  Spin-$s$ spherical harmonics and $\delta$,  {\em J. Math. Phys.} {\bf 8},  2155 (1967).

\bibitem{Plebanski1976}
J.~F. Pleba\'nski and M.~Demia\'nski,
    Rotating, charged and uniformly accelerating mass in general relativity,
    {\em Ann. Phys. (N.Y.)} {\bf 98}, 98 (1976).

\bibitem{Debever1971}
R.~Debever,  On type D expanding solutions of Einstein-Maxwell equations,
  {\em Bull. Soc. Math. Belg.} {\bf 23},  360 (1971).

\bibitem{Griffiths2005}
J.~B. Griffiths and J.~Podolsk{\'{y}},  Accelerating and rotating black
  holes,  {\em Class. Quantum Grav.} {\bf 22},  3467 (2005).

\bibitem{GriffithsPodolsky2006}
J.~B. Griffiths and J.~Podolsk\'y, A new look at the Pleba\'nski-Demia\'nki
  family of solutions,  {\em Int. J. Mod. Phys.~D} {\bf 15},  335 (2006).

\bibitem{PodolskyGriffiths2006}
J.~Podolsk\'y and J.~B. Griffiths,  Accelerating Kerr-Newman black holes in
  (anti-)de Sitter space-time,  {\em Phys. Rev.~D} {\bf 73}, 044018 (2006).

\bibitem{Vratny2018}
A.~Vr\'atn\'y, {\em Spacetimes with accelerating sources}.
Master Thesis, Faculty of Mathematics and Physics, Charles University, Prague, 2018.

\bibitem{PodolskyVratny2020}
J.~Podolsk\'y and A.~Vr\'atn\'y,  Accelerating NUT black holes,  {\em
  Phys. Rev.~D} {\bf 102}, 084024 (2020).

\bibitem{Matejov-2018}
D.~Matejov, {\em Twistor equation on isolated horizons}.
Master Thesis, Faculty of Mathematics and Physics, Charles University, Prague, 2018.

\bibitem{ChngMannStelea2006}
B.~Chng, R.~Mann, and C.~Stelea,  Accelerating Taub-NUT and Eguchi-Hanson
  solitons in four dimensions,  {\em Phys. Rev.~D} {\bf 74}, 084031 (2006).

\bibitem{BukLewandowski}
E.~Buk and J.~Lewandowski, Axisymmetric, extremal horizons at the presence of cosmological constant, (2020) arXiv:2012.15655 [gr-qc].

\bibitem{ChruscielCosta}
P.~T.~Chrusciel and J.~Lopes~Costa, On uniqueness of stationary vacuum black
holes, {\em Astérisque} {\bf 321}, 195 (2008).

\bibitem{HollandsIshibashi}
S.~Hollands and A.~Ishibashi, On the ‘Stationary Implies Axisymmetric’ Theorem
for Extremal Black Holes in Higher Dimensions, {\em Commun. Math. Phys.} {\bf 291}, 403 (2009).

\bibitem{Lewandowski-Szereszewski}
J.~Lewandowski and A.~Szereszewski, Axial symmetry of Kerr spacetime without the rigidity theorem,
{\em Phys. Rev.~D} {\bf 97}, 124067 (2018).

\bibitem{Dobkowski-Lewandowski-Pawlowski-2018}
D.~Dobkowski-Rylko, J.~Lewandowski, and T.~Pawlowski, The Petrov type D isolated null surfaces
{\em Class. Quantum Grav.} {\bf 35}, 175016 (2018).

\bibitem{Lewandowski-Ossowski}
J.~Lewandowski and M.~Ossowski, Non-singular Kerr-NUT-de~Sitter spacetimes,
{\em Class. Quantum Grav.} {\bf 37}, 205007 (2020).

\bibitem{Lewandowski-OssowskiII}
J.~Lewandowski and M.~Ossowski, Projectively non-singular horizons in Kerr-NUT-de Sitter spacetimes, {\em Phys. Rev. D} {\bf 102}, 124055 (2020).

\end{thebibliography}
\end{document}